\documentclass[10pt,a4paper]{article}

\usepackage{arxiv}
\usepackage{amsmath, amssymb, amsfonts}
\usepackage[utf8]{inputenc} 
\usepackage[T1]{fontenc}    

\usepackage{hyperref}       
\hypersetup{
    colorlinks=true,
    citecolor = blue,
    linkcolor=blue,
    filecolor=blue,      
    urlcolor=black,
}

\usepackage{xr}
\usepackage[capitalise]{cleveref}

\usepackage{newtxtext}
\usepackage{datetime2}
\usepackage{graphicx}
\usepackage{longtable}
\usepackage{multirow}
\usepackage{algorithm, algorithmic} 
\usepackage[nocompress]{cite}
\usepackage{todonotes}
\usepackage{url}
\usepackage{color, soul}

\usepackage{booktabs}
\usepackage{longtable}
\usepackage{array}
\usepackage{multirow}
\usepackage{wrapfig}
\usepackage{float}
\usepackage{colortbl}
\usepackage{pdflscape}
\usepackage{tabu}
\usepackage{threeparttable}
\usepackage{threeparttablex}
\usepackage[normalem]{ulem}
\usepackage{makecell}
\usepackage{xcolor}
\usepackage{tabularx}

\usepackage{setspace}
\usepackage{caption}

\usepackage{subcaption}

\usepackage{multirow}
\usepackage{booktabs}
\usepackage{adjustbox}

\newcommand{\etal}{\textit{et al.}\ }
\newcommand{\pag}{Pagendam \etal~\cite{pagendam_modelling_2020}}

\newcommand{\alb}{\textit{Ae.\ albopictus}}
\newcommand{\wpip}{\textit{w}Pip}
\newcommand{\arwp}{AR\textit{w}P}
\newcommand{\walba}{\textit{w}AlbA}
\newcommand{\walbb}{\textit{w}AlbB}
\newcommand{\walbab}{\textit{w}AlbAB}

\emergencystretch=\maxdimen
\title{Modelling \textit{Aedes albopictus} management, incorporating immigration and bidirectional \textit{Wolbachia} interactions}

\author{M. Ryan, M. Mendiolar, D. Pagendam, R. I. Hickson, and B. Trewin}

\author{
\href{https://orcid.org/0000-0003-2373-4384}{\includegraphics[scale=0.06]{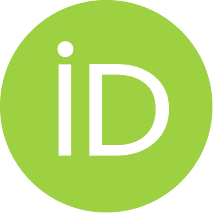}\hspace{1mm}Matthew Ryan} \\
	Commonwealth Scientific and Industrial Research Organisation (CSIRO) \\
	Adelaide, Australia \\
    	\AND
	\href{https://orcid.org/0000-0002-8380-819X}{\includegraphics[scale=0.06]{orcid.pdf}\hspace{1mm}Manuela Mendiolar} \\
    Commonwealth Scientific and Industrial Research Organisation (CSIRO) \\
	Dutton Park, Australia \\
    	\AND
	\href{https://orcid.org/0000-0002-8347-4767}{\includegraphics[scale=0.06]{orcid.pdf}\hspace{1mm}Dan Pagendam} \\
    Commonwealth Scientific and Industrial Research Organisation (CSIRO) \\
	Dutton Park, Australia \\
	\AND
	\href{https://orcid.org/0000-0001-6453-7745}{\includegraphics[scale=0.06]{orcid.pdf}\hspace{1mm}Roslyn I. Hickson} \\
	James Cook University and \\
    Commonwealth Scientific and Industrial Research Organisation (CSIRO) \\
	Townsville, Australia \\
    	\AND
	\href{https://orcid.org/0000-0002-7244-4044}{\includegraphics[scale=0.06]{orcid.pdf}\hspace{1mm}Brendan Trewin} \\
    Commonwealth Scientific and Industrial Research Organisation (CSIRO) \\
	Dutton Park, Australia \\
}

\date{}

\renewcommand{\shorttitle}{Modelling \textit{Aedes albopictus} management}

\begin{document}

\maketitle








\begin{abstract}
    \textit{Aedes albopictus} mosquitoes are competent vectors for the spread of at least 24 different arboviruses, including dengue, Ross River, and Japanese encephalitis viruses. However, they remain less studied than their more urban cousins, \textit{Aedes aegypti}. We model an Incompatible Insect Technique (IIT) strategy for mosquito control, with bi-directional incompatibility between two strains of \textit{Wolbachia} (\walba/\walbb\, $\times$ \arwp) and age-based cytoplasmic incompatibility decay in a well-mixed population.  An important consideration in bi-directional IIT control programs is reversibility when immigration is included.  We explore the establishment probability after female contamination of an artificially-infected \textit{Wolbachia} mosquito strain, finding a conservative threshold of 40\% likely driven by mating inefficiencies -- this threshold needs validation in future field and lab experiments. We consider the suppression dynamics and probability of mosquito management success for different release strategies, showing differences in success between release cessation  and six months later for different immigration rates. Importantly, our model suggests bi-directional IIT control programs are reversible with low amounts of immigration. We determine a corresponding cost proxy (numbers of mosquitoes released), showing similar short-term costs with differences in medium- and longer-term costs between release strategies. This work demonstrates opportunities to optimise the suppression of these medically important mosquitoes.
\end{abstract}


\section{Introduction}

\textit{Aedes albopictus} is the most invasive mosquito species in the world \cite{GISD_2024, benedict_spread_2007}. This species is capable of transmitting many pathogens to humans including chikungunya, Zika and Japanese encephalitis \cite{zhang_review_2023, leta_global_2018}, and is responsible for spreading dengue into areas where the disease had once been eliminated \cite{cochet_autochthonous_2022, quam_dissecting_2016,  zeng_standalone_2022}. Most control programs for \alb\ rely on traditional approaches, including removal of breeding sites in urban areas and the application of chemical insecticides \cite{fonseca_area_2013}. Resistance to these chemical interventions is increasing \cite{auteri_insecticide_2018, asgarian_worldwide_2023}, while removal of larval habitat is labour-intensive and inefficient, as it does not remove cryptic or inaccessible sites \cite{kay2000importance}. As such, new and innovative population control strategies that scale across large urban environments are urgently needed. 

\textit{Wolbachia}-based population suppression and replacement strategies are well-established for the control of \textit{Ae. aegypti} mosquitoes and the diseases they spread (see, for example, \cite{crawford_efficient_2020, zheng_incompatible_2019, hoffmann_successful_2011, nazni_establishment_2019, beebe_releasing_2021}). Cytoplasmic incompatibility (CI), a biological trait of the endosymbiotic bacteria \textit{Wolbachia}, is an important component of both suppression \cite{beebe_releasing_2021, crawford_efficient_2020, zheng_incompatible_2019} and replacement control strategies \cite{hoffmann_successful_2011}. This natural mechanism of reproductive manipulation by \textit{Wolbachia} on its insect host phenotype is referred to as a uni-directional (one-way) CI and alters the outcomes of mating.  Only crosses between both \textit{Wolbachia} infected male and female mosquitoes, or uninfected wild-type males and \textit{Wolbachia} infected females,  produce viable offspring in the next generation \cite{barr_cytoplasmic_1980, laven_eradication_1967}.  The Incompatible Insect Technique (IIT) is a management technique which utilises CI, taking advantage of the uni-directional incompatibility exhibited by \textit{Wolbachia}-infected males and a wild population that is uninfected \cite{boller_evidence_1974, boller_incompatible_1976}. Research trialling \textit{Wolbachia} as a control strategy for \textit{Ae. aegypti} have recently achieved successful suppression of populations in field trials \cite{crawford_efficient_2020, zheng_incompatible_2019, beebe_releasing_2021}. While the \textit{Ae. aegypti} uni-directional CI system is well studied, suppressing populations of other medically important mosquito species requires further investigation. 

There has been substantially less focus on the control of \alb\ via the IIT. 
\alb\ exhibits different fitness characteristics when compared to \textit{Ae. aegypti}, such as reproduction rate, average life expectancy and an ability to overwinter in colder regions. Furthermore, cage experiments and modelling of \textit{Ae. aegypti} systems to identify uni-directional invasion thresholds for \textit{Wolbachia} establishment (see, for example, \cite{laporta_global_2023, pagendam_modelling_2020}) are not directly applicable to \alb\ as wild populations of the species carry \walba/\walbb\ strains of \textit{Wolbachia} and exhibit a natural form of CI expressed through the \walba\ strain \cite{ogunlade2022modelling}.

In recent years, a number of artificial \textit{Wolbachia} infections have been introduced into \alb, including single \cite{moretti_male_2013} and triple infections of the Pipientis \textit{Wolbachia} (\wpip) from \textit{Culex} mosquitoes \cite{zhang_combining_2015}. In particular, the injection of \alb\ with the \arwp\ form of \wpip\ has been successfully deployed in IIT field trials \cite{caputo2020bacterium, mains_female_2016}.
When two incompatible \textit{Wolbachia} infected mosquitoes mate, all subsequent offspring of these matings do not survive. This process is referred to as bi-directional CI \cite{moretti_cytoplasmic_2018}. Control programs applying a bi-directional CI strategy introduce several unknown factors impacting management decision-making, including the level of population suppression that can be achieved as influenced by \textit{Wolbachia} invasion thresholds, particularly if \textit{Wolbachia}-infected females are released through imperfect sex-separation \cite{pagendam_modelling_2020, lombardi_incompatible_2024}.

Despite increasing improvements to mosquito sex-separation technologies \cite{crawford_efficient_2020} some methods are imperfect and have the potential to release artificially-infected \textit{Wolbachia} females into the landscape, risking the undesired outcome of population replacement \cite{project_wolbachia_2021, ross_designing_2021}. Typical sex-separation technologies using sieves result in a small female contamination rate of around $1-5$\% \cite{project_wolbachia_2021}. Establishing \textit{Wolbachia} in an \textit{Ae. aegypti} IIT control program at a threshold of greater than $20\%$ should lead to a population replacement outcome \cite{zhang_combining_2015, project_wolbachia_2021}, as wild populations have historically not contained \textit{Wolbachia}.  In this case, population replacement is an undesirable outcome of an IIT control program as the released \textit{Wolbachia} would no longer be effective at achieving its goal of suppression \cite{dobson2002effect}.

Typically, a combination of both IIT and a radiation treatment (Sterile Insect Technique) have been used as a method to prevent unwanted establishment of an artificial \textit{Wolbachia} strain \cite{soh_strategies_2022, zheng_incompatible_2019}. However, natural \alb\ populations containing \walba/\walbb\ exhibit CI and could theoretically suppress the establishment of an artificial \textit{Wolbachia} strain, should females be released \cite{moretti_cytoplasmic_2018}. This opens the potential for management decisions which prevent establishment and continue suppression of the wild population with imperfect sex separation. 

While establishing an artificial \textit{Wolbachia} strain in a wild \walba/\walbb\ population may be minimised, this is not guaranteed, as the \walba\ concentration in wild \alb\ males decreases as they age. This reduces the effectiveness of \walba\ CI in \alb\ from $100$\% at $1-15$ days to $68$\% at $15-19$ days, and $0$\% (i.e. no incompatibility) at $20$ days and beyond \cite{calvitti_wolbachia_2015}. This decrease imparts a relative reproductive advantage to the \arwp\ \alb\ population used in the IIT control program, as \arwp\ infected females can successfully mate with older wild-type males, while \arwp\ mosquitoes also exhibit marginally improved fitness to the wild \walba/\walbb\ population \cite{calvitti_wolbachia_2015}.

Thus, investigating IIT control programs for the management of \alb\ presents interesting and unique challenges. The combination of different fitness parameters, bi-directional CI, and age-related decay in CI means that previous studies for \textit{Ae. aegypti} are not immediately applicable to \alb. Further, a challenge for the regulation of these types of new technologies is the risk an IIT control program leads to unintended consequences, such as a population replacement rather than suppression.  Given imperfect sex-separation is common, there is a genuine concern of establishing an artificial \textit{Wolbachia} stain when applying the IIT.  This risk may be mitigated by natural immigration and emigration of wild \alb\ populations from surrounding areas.  To investigate the complex nature of a bi-directional IIT control program for \alb\ while considering the impacts of immigration, we use mathematical modelling as a cost-effective tool to explore a range of population suppression scenarios.

Mathematical and computational models have been used to explore the impacts of multiple \textit{Wolbachia} strains as a control agent for mosquito populations.  These models typically focus on either understanding conditions of co-existence of multiple \textit{Wolbachia} strains in the wild \cite{keeling2003invasion, ogunlade2022modelling, farkas2010structured} or the effects of bi-directional CI on IIT control programs \cite{dobson2002effect, moretti_cytoplasmic_2018, turelli2010cytoplasmic, soh_strategies_2022}.  In particular, Moretti \textit{et al.} \cite{moretti_cytoplasmic_2018} built upon the discrete generation model of Dobson \textit{et al.} \cite{dobson2002effect} to understand how the introduction of \arwp\ \textit{Wolbachia} may be used to control wild-type \alb\ populations.  In their approach, Moretti \textit{et al.} used a deterministic model to investigate non-overlapping generations of mosquitoes. However, deterministic models can have significant limitations for small population sizes where small stochastic changes can have a large impact on the dynamics \cite{pagendam_modelling_2020}, and discrete generations precludes the modelling of the age-based CI decay found in wild-type \alb\ mentioned above.

Here we build on an existing stochastic computational model for \textit{Ae. aegypti} developed by \pag\ to explore suppression scenarios applying bi-directional CI and \arwp\ \textit{Wolbachia} establishment thresholds on a theoretical population of \alb. Using the extended model, we focus on hypothetical scenarios where \arwp\ infected \alb\ are introduced into a background of wild-type \walba /\walbb. We explore scenarios where populations of wild \alb\ are managed without uncontrolled replacement by an established \arwp\ population and we determine how key differences in suppression affect management success, including a proxy for cost.

\section{Methods}

When constructing our model, we are motivated by modelling a single urban block of houses surrounded by many un-modelled patches \cite{manica2016spatial, pagendam_modelling_2020}.  From a mathematical perspective, this means we restrict our investigations to a single patch model.  Our model can be extended to patches larger than an urban block (for example, an isolated island) by modifying the parameter estimates and initial conditions of the model. A description of all model parameters and their estimates used are found in Supplementary Table \ref{supp-tab:parameters}, with the stoichiometries fully described in Table \ref{supp-tab:transitions}.

\subsection{Model description}

\begin{figure}[htbp]
    \centering
    \includegraphics[width=1\linewidth]{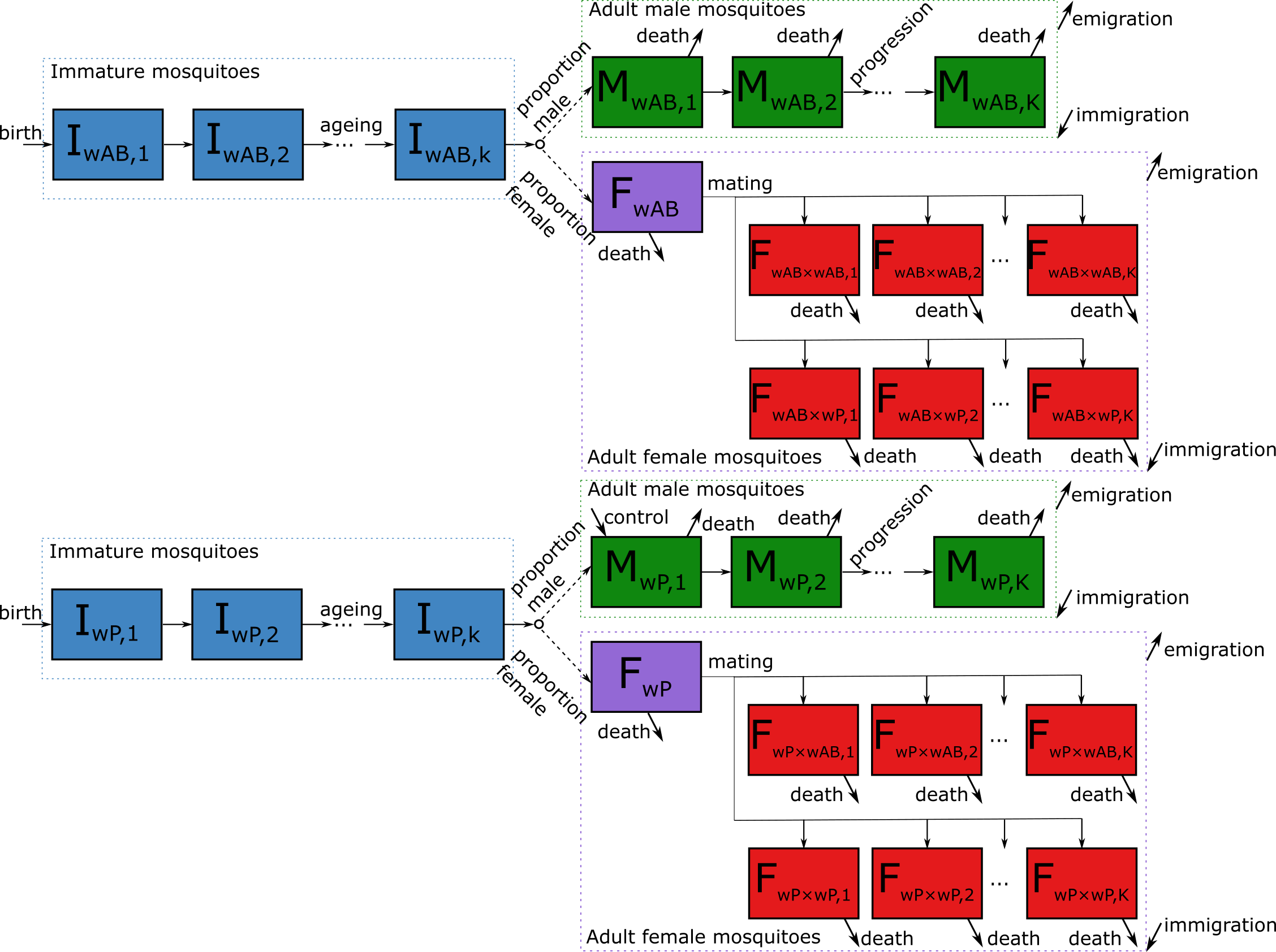}
    \caption{Schematic of the mosquito population model, showing progression from the immature stages (blue) through to adult males (green) or females who are unmated (purple). Females then progress to those mated with males with a specific \textit{Wolbachia} type and Cytoplasmic Incompatibility efficacy (red). Immigration and emigration occurs from each adult compartment, captured here by the flows from the dashed boxes grouping compartments, though in the scenarios we study there is no immigration of \arwp\ mosquitoes. For readability, \walbab\  has been abbreviated to wAB and \arwp\ has been abbreviated as wP. In the $F_{wX \times wJ}$ notation the first strain ($wX$) refers to the \textit{Wolbachia} strain the female carries, and the second ($wJ$) to the male.}
    \label{fig:compartmental_diagram}
\end{figure}

\pag\ proposed a \textit{Wolbachia}-IIT Markov Population Process model to investigate the efficacy of different release strategies on the risk of \textit{Wolbachia} establishment for \textit{Ae. aegypti} populations.  We extend their model to \alb\ and include age-related decay in \textit{Wolbachia} CI, as well as immigration and emigration for wild adult mosquito populations (Figure \ref{fig:compartmental_diagram}). Many components of our model correspond to the \pag\ model, which we recap briefly for completeness.  For brevity throughout, we refer to \alb\ mosquitoes infected with either \walba/\walbb\ as \walbab.

Pagendam \textit{et al.}'s model proposes that there is a pool of future adult mosquitoes, or ``immature'' mosquitoes, that spend a Gamma-distributed amount of time as immatures.  Immature  mosquitoes include the eggs, larvae, and pupae in the mosquito population that will survive to adulthood. When immatures age to become adults, there is a $50$\% chance they will either become adult males or unmated adult females.  Adult males can then mate with unmated females, and we assume that females only mate with one male in their lifespan.  Male and female mosquitoes mate at a species-dependent rate, modified by Fried's index. That is, if there are more male mosquitoes of a particular type (\walbab\ or \arwp) then females are more likely to mate with that type of male.  Mated female adults then give birth to immatures at a rate $\tilde{\lambda}$. The birth rate of immatures is modified by a density-dependent rate that controls for the size of the immature pool; this emulates the density dependence of higher larval mortality at higher larval population densities.  Note that both \walbab\ and \arwp\ immature mosquitoes contribute to the same immature pool and hence the density-dependent rate.

Our proposed model considers a bi-directional CI approach where mated females $F_{\walbab\times \arwp}$ and $F_{\arwp\times \walbab}$ fail to produce viable offspring, where the first subscript denotes the \textit{Wolbachia} strain of the female mosquito, and the second subscript indicates the strain of the male. However, CI decays with the age of male \walbab\ mosquitoes, meaning that as male \walbab\ mosquitoes age there is a higher chance for $F_{\arwp\times \walbab}$ to produce viable \arwp\ offspring. To capture this age-related decay of CI, we discretise the male mosquitoes into $K$ compartments, depicted in Figure \ref{fig:compartmental_diagram}, where $K=20$ in our study.  We use the CI values from Calvitti \textit{et al.} \cite{calvitti_wolbachia_2015} where the CI is $100\%$ effective for \walbab\ males in $M_{\walbab, 1}$--$M_{\walbab, 14}$ (that is, no viable offspring), $67\%$ effective for \walbab\  males in $M_{\walbab, 15}$--$M_{\walbab, 19}$, and $0\%$ effective for \walbab\ males in $M_{\walbab, 20}$ (that is, no CI preventing offspring). Along with increased fitness \cite{moretti_male_2013}, this gives the \arwp\ mosquitoes a slight reproductive advantage over the wild-type \walbab\ mosquitoes. It is important to note that the compartments $M_{w, j}$, for \textit{Wolbachia} strain $w$ and $j=1, 2,\dots, K$, do not capture the age of the mosquito, but rather the distribution of time a mosquito will spend with a given level of CI. That is, in our model adult mosquitoes have an exponentially distributed lifespan, but a truncated gamma distributed time spent with a given CI efficacy.

Another feature of our model is the inclusion of immigration and emigration.  Although we are only investigating a single-patch model, the patch represents a city block of urban houses. City blocks do not exist in isolation, and this may have consequential effects on any suppression control programs being examined. 
We assume that all neighbouring patches, not explicitly modelled, are in an equilibrium steady state for the wild-type \walbab\  mosquitoes, so all immigrant mosquitoes are of the wild-type.  However, we allow both the wild-type and the invasive \arwp\ to emigrate out of the modelled patch.  Immigration and emigration rates are calculated such that they leave a net-zero change in the steady state wild-type population; see the Supplementary Material for more details.

Our model is biologically feasible when $\frac{\mu_F}{\frac{1}{2} \lambda p_{mated}} < 1$, where $\mu_F$ is the female death rate, $\lambda$ is the base birth rate of future adult mosquitoes, and $p_{mated}$ is the proportion of wild-type female mosquitoes that are mated in the steady state equilibrium; see the Supplementary Material for the derivation of this constraint.  This constraint means that the observed female death rates from previous studies \cite{vavassori_active_2019} are not feasible under this model.  As such, in all investigations in this paper we have derived the female death rate ($\mu_F$) as the largest biologically feasible value satisfying $\frac{\mu_F}{\frac{1}{2} \lambda p_{mated}}=0.999$ for fixed $p_{mated}$ and $\lambda$.

A generalised version of this model for $J$ strains of \textit{Wolbachia} is presented in the Supplementary Materials.

\subsection{Simulation studies}

We perform two separate simulation studies.  The first investigates the unstable equilibrium threshold for when the \arwp\ infected \alb\ are likely to establish in the wild-type population.  The second investigates the effects of immigration and emigration for three different IIT release strategies of \arwp\ infected \alb\  into the wild-type population.

\subsubsection{Cage simulations}

A major concern with IIT control programs is the establishment of the artificial \textit{Wolbachia} strain in the wild mosquito population, in this case \alb\ infected with \arwp.  To mitigate this concern, IIT control programs aim to keep the artificial strain under an unstable equilibrium threshold (or probability) of establishment, which we denote by $\omega^*$.  The value of $\omega^*$ is defined such that if the proportion of the artificial \textit{Wolbachia} strain (\alb\ infected with \arwp) relative to the initial wild-type population (\alb\ infected with \walbab) exceeds $\omega^*$, then the artificial strain is more likely to establish.

The unstable equilibrium threshold of establishment $\omega^*$ has been studied for \textit{Ae. aegypti} in the context of a uni-directional CI IIT control program \cite{axford_fitness_2016}.  Axford \etal found through cage experiments that $\omega^*$ is between $0.20$ and $0.25$ for \textit{Ae. aegypti} when the invasive \textit{Ae. aegypti} are infected with \walbab\ \cite{axford_fitness_2016}, which has been validated \textit{in silico} \cite{pagendam_modelling_2020}. However, 
there is no \textit{a priori} reason to expect that $\omega^*$ will remain unchanged in a bi-directional CI IIT control program.  In fact, Moretti \textit{et al.} \cite{moretti_cytoplasmic_2018} suggest that the unstable equilibrium threshold is inappropriate for \alb.  Further, Lombardi \textit{et al.} \cite{lombardi_incompatible_2024} recently conducted cage experiments for \alb\ to consider the effects of bi-directional CI on IIT control programs and found an unstable equilibrium of approximately $\omega^* = 0.40$ (although some of these experiments resulted in establishment).  We aim to investigate these results further using \textit{in silico} cage experiments.

Here, we investigate feasible values for $\omega^*$ in the context of a bi-directional CI IIT control program for \alb\ using \textit{in silico} cage simulations. As qualitative controls, we also investigate values of $\omega^*$ when CI is uni-directional and when CI is bi-directional but without age-related decay.  Our primary scenario starts with an initial \alb\ adult mosquito population of $420$, equally split between males and females, and varies the initial proportion of mosquitoes infected with \arwp\ from $0.05$ to $0.50$ by increments of $0.05$.  As a qualitative control for these cage simulations, we compare our uni-directional results with those from the \pag\ model, which were in turn directly compared with experimental findings. These qualitative control simulations are equivalent to a wild-type \alb\ population which has been cleared of \textit{Wolbachia}. For each initial proportion of \arwp, we run $1\ 000$ simulations.  

Simulations were run for up to $500$ simulation days or for $180$ days after one of the following two stopping conditions were met: 1) the number of \arwp\ adults reached $0$ or 2) the adult \alb\ (either \walbab\ or cleared) population was suppressed, where suppression is defined as the population dropping below $10\%$ of the initial population 
(that is, when the number of adult \alb\  mosquitoes dropped below $42$). A $10\%$ initial population criteria is more conservative than the  $30\%$ population suppression threshold used by Fonseca \textit{et al.} \cite{fonseca_area_2013}. The cage simulations were run for the expected fitness parameters, as well as low and high fitness parameters (see Supplementary Materials Table~\ref{supp-tab:parameters}).

\subsubsection{Management decisions:  IIT release scenarios} 

We examined three different IIT release scenarios and the effects of different immigration and emigration rates on the efficacy of these control programs. Each of the IIT release scenarios followed the same basic principles, but differed on when the release was terminated. The basic scenario was as follows: 
every seven days, \arwp\ mosquitoes will be released into the patch at an overflooding ratio of $5:1$, meaning that the number of \textit{Wolbachia}-infected males released into the population at each event was five times the current wild-type male population size. Release also assumed a female contamination rate of $0.01$ \cite{moretti_cytoplasmic_2018, lombardi_incompatible_2024} with
scenarios running for at least $100$ days before any stopping condition was considered. The three stopping conditions were:
\begin{enumerate}
    \item[] \textbf{Na\"ive scenario}: \textit{Wolbachia} release will run until the adult \walbab\ population drops below $10\%$ of the initial steady state population.
    \item[] \textbf{Complete stop scenario:} \textit{Wolbachia} release will run until the adult \walbab\ population drops below $10\%$ of the initial steady state population \textit{OR} the proportion of \arwp\ adults exceeds the unstable equilibrium threshold $\omega^*$.
    \item[] \textbf{Maintain scenario:} \textit{Wolbachia} release will run until the adult \walbab\ population drops below $10\%$ of the initial steady state population \textit{OR} the proportion of \arwp\ adults exceeds the unstable equilibrium threshold $\omega^*$.  Releases will resume if the proportion of \walbab\ adults returns to more than $10\%$ of the initial population (that is, they re-establish) \textbf{AND} the proportion of \arwp\ adults drops below $0.8 \omega^*$.
\end{enumerate}
The na\"ive scenario represents the simplest and most obvious baseline to achieve suppression of the wild-type population.  The complete stop scenario explores a cautious implementation of a dual objective scenario of suppressing the \walbab\ mosquito and preventing the artificial \textit{Wolbachia} strain from establishing by stopping releases at the conservative establishment threshold $\omega^*$.  The maintain scenario explores achieving these dual objectives, whilst continuing the release of \arwp\ under sustainable conditions (avoiding establishment: the second objective). Note that the condition for restarting releases when the proportion of \arwp\ drops below $0.8\omega^*$ was chosen for convenience, to prevent stop-start oscillations from having identical threshold values.

To investigate the extent that immigration and emigration will affect these release scenarios, we consider three mosquito immigration rates.
The rates chosen represent i) a closed, isolated population with no immigration, ii) a low immigration/emigration rate (2 adults per week), and iii) a high immigration/emigration rate (10 adults per week). 

For each release scenario and immigration/emigration rate, we ran $1\,000$ simulations. Simulations were run for $920$ simulation days ($2.5$ years), with \arwp\ releases occurring until the stopping condition was met or for a maximum of $730$ days ($2$ years). We ran simulations for six months after the cessation of all release scenarios to investigate the reversibility of the IIT control program.
At the end of the simulations, we determined the proportion of successful scenarios defined as suppressing the \walbab\ population below $10\%$ of the initial population size while keeping the invasive \arwp\ below the unstable equilibrium threshold $\omega^*$. We compared the proportion of successful scenarios at the end of the release strategy and again six months after to assess the scenario's longevity and reversibility. Additionally, we calculated a proxy for the cost of the release scenario based on the number of adult mosquitoes released during the lifetime of the simulation.  Simulations were run using expected, low, and high population parameters to explore sensitivity of results to possible extreme values (see Supplementary Materials Table~\ref{supp-tab:parameters}).

\section{Results}

\subsection{Cage simulations}

\begin{figure}[htbp]
    \centering
    \includegraphics[width=\linewidth]{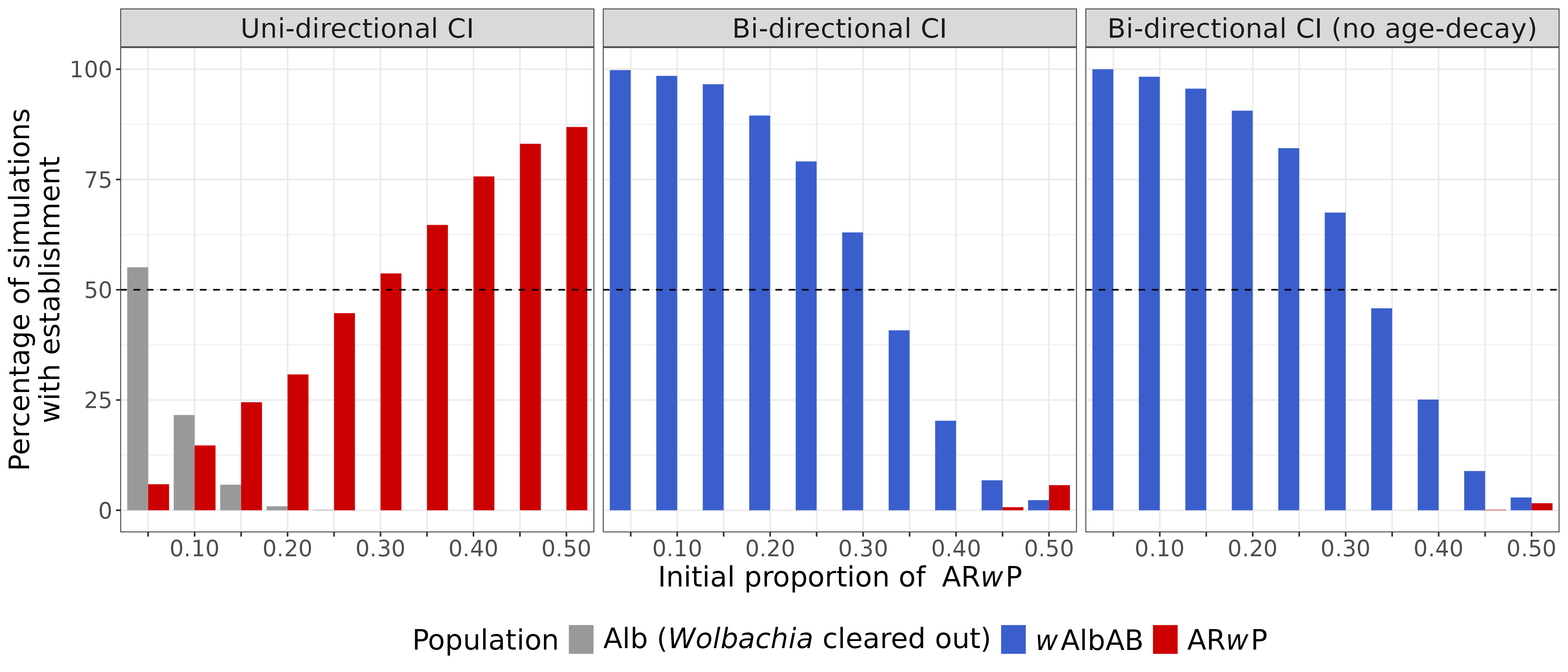}
    \caption{Percentage of \textit{in silico} cage simulations ending in population establishment at the end of the $500$ iterations. An \alb\ population is considered established if the proportion of adults is greater than $10\%$ of the initial population size. The left panel shows simulations with uni-directional Cytoplasmic Incompatibility (CI), the middle panel depicts bi-directional CI with age-decay and the right panel bi-directional CI without age-decay. Red columns represent the \arwp\ population, blue columns the \walbab\ population, and grey columns the \textit{Wolbachia} cleared (does not contain \walbab) \alb\ population. The $x$-axis indicates the initial proportion of \arwp, and the $y$-axis shows the percentage of simulations that resulted in establishment. The dotted horizontal line marks $50\%$ of simulations. Each simulation started with $420$ adult mosquitoes (half male, half female), and the fitness parameters are the expected values defined in Supplementary Materials Table~\ref{supp-tab:parameters}.
    }
    \label{fig:cage_establish}
\end{figure}

We found a higher unstable equilibrium threshold for \alb\ than \textit{Ae. aegypti} when only considering uni-directional CI \cite{pagendam_modelling_2020, axford_fitness_2016}.  
That is, \pag\ found for \textit{Ae. aegypti} that when the frequency of adults with a \textit{Wolbachia} infection starts at $25\%$ of the initial population the \textit{Wolbachia} strain will establish in at least $50\%$ of the simulations; for \alb\ we found the invasive strain needs to start at $30\%$ of the initial population to have the same property (where the red columns meet the horizontal dashed line, left panel, Figure \ref{fig:cage_establish}).

In contrast, the results for the bi-directional CI case are striking (middle panel, Figure \ref{fig:cage_establish}).  Our simulations showed that at least $45\%$ of the initial population is required to start as \arwp\ for simulations to result in an outcome where the artificial \textit{Wolbachia} strain established (red columns, middle panel, Figure \ref{fig:cage_establish}).  This bi-directional CI result is driven by \arwp\ failing to find compatible mates for smaller population sizes and are consequently driven to extinction by density effects.  Based on these results, we consider a conservative unstable equilibrium threshold as $\omega^* = 0.4$, the cut-off to prevent \arwp\ from becoming established in the wild. These results are qualitatively consistent for the high and low fitness parameters (Figures~\ref{supp-fig:cage_establish_low}-\ref{supp-fig:cage_establish_high}).

When removing age-related decay of CI in the bi-directional simulations, the same conservative threshold $\omega^*=0.4$ is suggested (Figure \ref{fig:cage_establish}, right panel).  However, there are qualitative differences in the likelihood of \arwp\ establishment across the two bi-directional scenarios.  When CI decays with mosquito age, \arwp\ is more likely to establish in the \textit{in silico} cage population for initial proportions of \arwp\ greater than $0.4$. The effect of age-related decay is more pronounced for the extreme fitness parameters, particularly at the $50\%$ relative abundance threshold (Figures~\ref{supp-fig:cage_establish_low}-\ref{supp-fig:cage_establish_high}).

\subsection{Comparison of IIT release scenarios}

\begin{figure}[htbp]
    \centering
    \includegraphics[width=\linewidth]{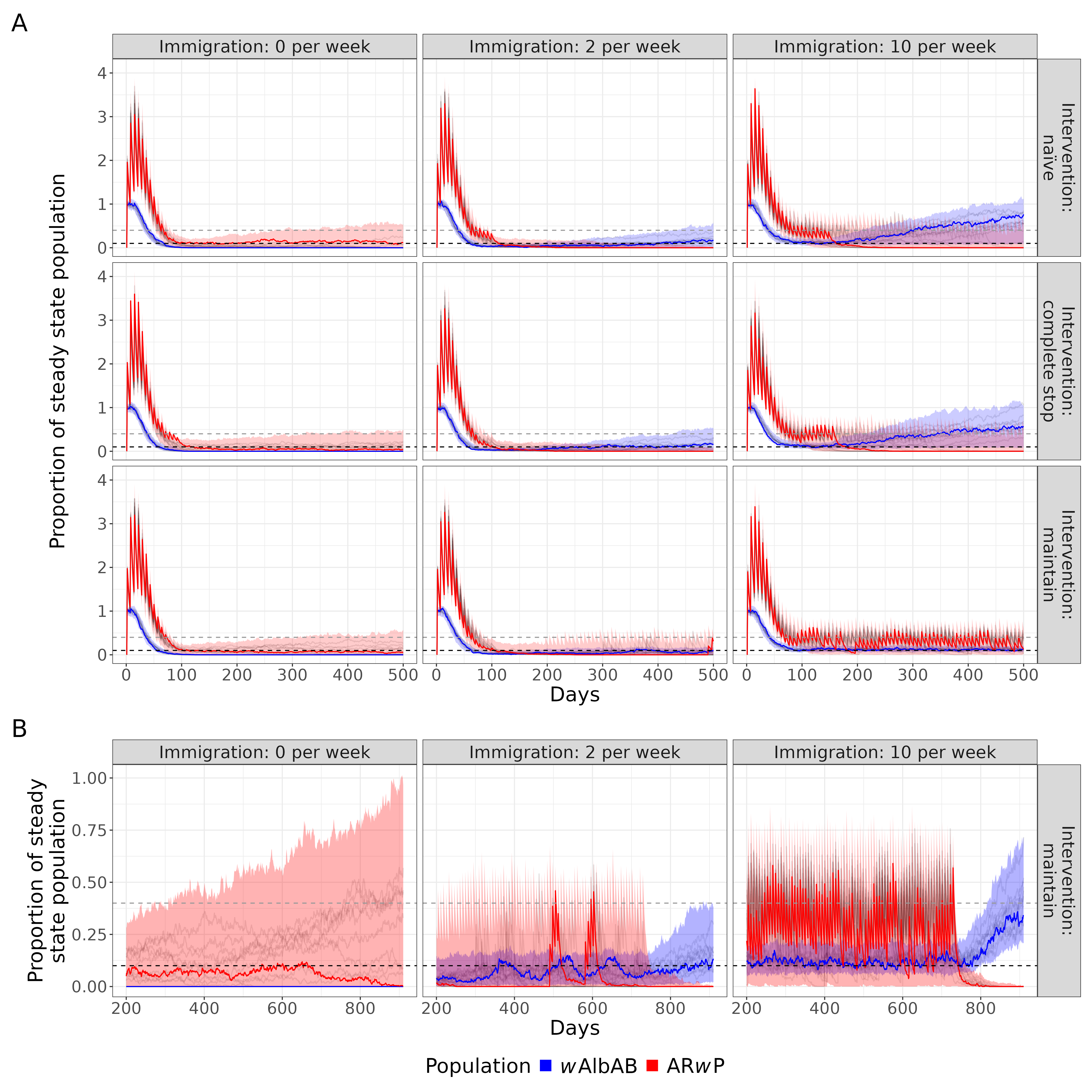} 
    \caption{Proportion of mosquitoes relative to the steady-state population over time for different release scenarios (rows) and immigration rates (columns). The \arwp\ population (red) and \walbab\ population (blue) lines represent the median with $95\%$ confidence intervals (shaded areas). Black lines represent $10$ random simulation runs. Panel A displays the population over $500$ days under three immigration rates ($0$, $2$, and $10$ mosquitoes per week). The intervention scenarios include na\"ive, complete, and maintain. Panel B focuses on the maintain scenario over days 200 to 920 for the same immigration rates. Horizontal dotted lines represent the suppression/establishment ($10\%$) and the unstable equilibrium ($40\%$) thresholds used. Values above $1$ on the y-axis indicate population levels higher than the initial steady state. Model parameters are the expected values  defined in Supplementary Materials Table~\ref{supp-tab:parameters}.}
    \label{fig:intervention_stormclouds}
\end{figure}

The maintain scenario was the suppression strategy most likely to reflect the control of a mosquito population using the IIT in the real-world. Here the rate of immigration and emigration are important as they increase the wild \alb\ population, influencing the decision to resume or maintain releases of \arwp\ infected males. There were stark contrasts between the  immigration rates under a maintain scenario (Figure \ref{fig:intervention_stormclouds} B). High immigration, when compared to  low immigration rates, influences the total number of released \arwp\ infected males required to suppress the wild population and also increases the total size of both established \textit{Wolbachia} populations.  This suggests it may be difficult to maintain suppression on a wild mosquito population with high immigration rates. In contrast, a low immigration rate ($2$ adults per week) leads to an extended period of low growth after suppression in the wild \alb\ population (Figure \ref{fig:intervention_stormclouds} A). In our low immigration, maintain scenario this occurred between approximately 120 and 220 days, after which \arwp\ releases restarted due to the wild \alb\ population increasing to $10\%$ of its original size. 

There is no qualitative difference between the na\"ive and complete stop release scenarios for all immigration rates (first two rows, Figure \ref{fig:intervention_stormclouds} A). The na\"ive scenario resulted in the recovery of the wild \walbab\ population after releases ceased under low and high immigration rates (Figure \ref{fig:intervention_stormclouds} A, top row). Likewise, wild-type populations recovered when \arwp\ releases ceased during the complete stop  scenarios with low and high immigration (Figure \ref{fig:intervention_stormclouds} A, middle row). Variation in the rate of wild \alb\ population recovery in both the naïve and complete stop scenarios was dependent upon total immigration/emigration, with high immigration rates leading to an almost complete rebound after $500$ days in the majority of simulations. Interestingly, under low immigration and both the na\"ive and complete stop scenarios, the wild population was less than $40\%$ of its initial population ($<168$ adult mosquitoes) for the majority of simulations after 500 days, indicating a slow rebound for the wild-type population. 

Low levels of \arwp\ replacement did occur across all release scenarios with no immigration, even after suppression was successful and releases ceased (Figure \ref{fig:intervention_stormclouds} A, first column). This is the result of low levels of female contamination introduced during the overflooding stage of the first $100$ days of scenarios.

These results are consistent across the high and low fitness parameters, with the caveat that there is no qualitative difference between the three strategies when immigration is low for the high fitness parameters (Figures~\ref{supp-fig:intervention_stormclouds_low} and \ref{supp-fig:intervention_stormclouds_high}).

When age-related decay of CI is removed from the bi-directional model, there are slight qualitative differences in the results (Figure \ref{supp-fig:intervention_costs_nodecay}).  Specifically, when immigration is high in the na\"ive scenario, the expected time to suppression of the wild-type \walbab\ population is longer, requiring more continued releases of the invasive \arwp\ strain on average.

\subsection{Success of IIT release scenarios}

The \arwp\ population is kept below $40\%$ of the initial population size for all IIT release scenarios and all immigration rates, both at the time release is stopped and six months after (Table~\ref{tab:intervention_establishment}), indicating successful outcomes.  Within six months of releases ceasing the \arwp\ population is kept below $168$ mosquitoes across all simulations, likely due to lack of viable mates.  For the \walbab\ population, IIT release scenarios are successful for all immigration rates and release scenarios, except for the maintain scenario when immigration is high (Table \ref{tab:intervention_establishment}).  Since releases are ongoing and immigration is high in this scenario, the \walbab\ population is kept above 42  in more than $88\%$ of simulations.  Six months after releases of \arwp\ mosquitoes are ceased, we observe evidence of reversibility of the IIT control program when immigration is low, and complete reversibility when immigration is high.

These results are qualitatively similar for the low fitness parameters (Table~\ref{supp-tab:intervention_establishment_low}). For the high fitness parameters, the only scenario that successfully suppresses the \walbab\ population and keeps the \arwp\ population below the unstable equilibrium threshold is the na\"ive scenario under no or low immigration rates.  However, for these parameters all scenarios are successful within six months of ceasing releases except when immigration is high, where IIT release scenarios are reversible for the wild-type \walbab\ population (Table~\ref{supp-tab:intervention_establishment_high}).  When age-decay of CI is removed from the model, there is a slight reduction in management success of the \walbab\ population after 6 months for low immigration rates (Supplementary Materials Table \ref{supp-tab:intervention_establishment_Expected_nodecay}).

\begin{table}[htbp]
    \caption{Percentage of simulations with successful outcomes for \walbab\ and \arwp\ populations under different immigration rates and release scenarios. A successful outcome is defined as keeping the \walbab\ population below $10\%$ of the initial population size, and \arwp\ below $40\%$ (unstable equilibrium threshold). The data is shown both within the stopping time and six months after.}
    \label{tab:intervention_establishment}
    \centering
    \begin{tabular}[t]{lrrrrrr}
    \toprule
    &  \multicolumn{2}{c}{Na\"ive} & \multicolumn{2}{c}{Complete Stop}  & \multicolumn{2}{c}{Maintain}\\
    
    \cmidrule(lr){2-3}
    \cmidrule(lr){4-5}
    \cmidrule(lr){6-7}
    Scenario
    & \walbab & \arwp & \walbab & \arwp & \walbab & \arwp \\
    \midrule
    
    \addlinespace[0.3em]
    \multicolumn{7}{l}{\textbf{0 mosquitoes per week}}\\
    \rowcolor{gray!6} {\hspace{1em}Within stopping time} & 100.0 & 100.0 & 100.0 & 100.0 & 100.0 & 100.0\\
    \hspace{1em}After six months & 100.0 & 99.9 & 100.0 & 100.0 & 100.0 & 100.0 \\

    \addlinespace[0.3em]
    \multicolumn{7}{l}{\textbf{2 mosquitoes per week}}\\
    \rowcolor{gray!6}{\hspace{1em}Within stopping time} & 100.0 & 100.0 & 100.0 & 100.0 & 92.2 & 100.0\\
    \hspace{1em}After six months & 81.8 & 100.0 & 80.1 & 100.0 & 52.2 & 100.0 \\
    
    \addlinespace[0.3em]
    \multicolumn{7}{l}{\textbf{10 mosquitoes per week}}\\
    \rowcolor{gray!6}{\hspace{1em}Within stopping time} & 99.8 & 100.0 & 99.7 & 99.7 & 21.2 & 100.0\\
    \hspace{1em}After six months & 0.0 & 100.0 & 0.0 & 100.0 & 0.0 & 100.0 \\
    \bottomrule
    \end{tabular}
\end{table}

\subsection{``Cost'' of an IIT control program}

\begin{figure}[htbp]
    \centering
    \includegraphics[width=\textwidth]{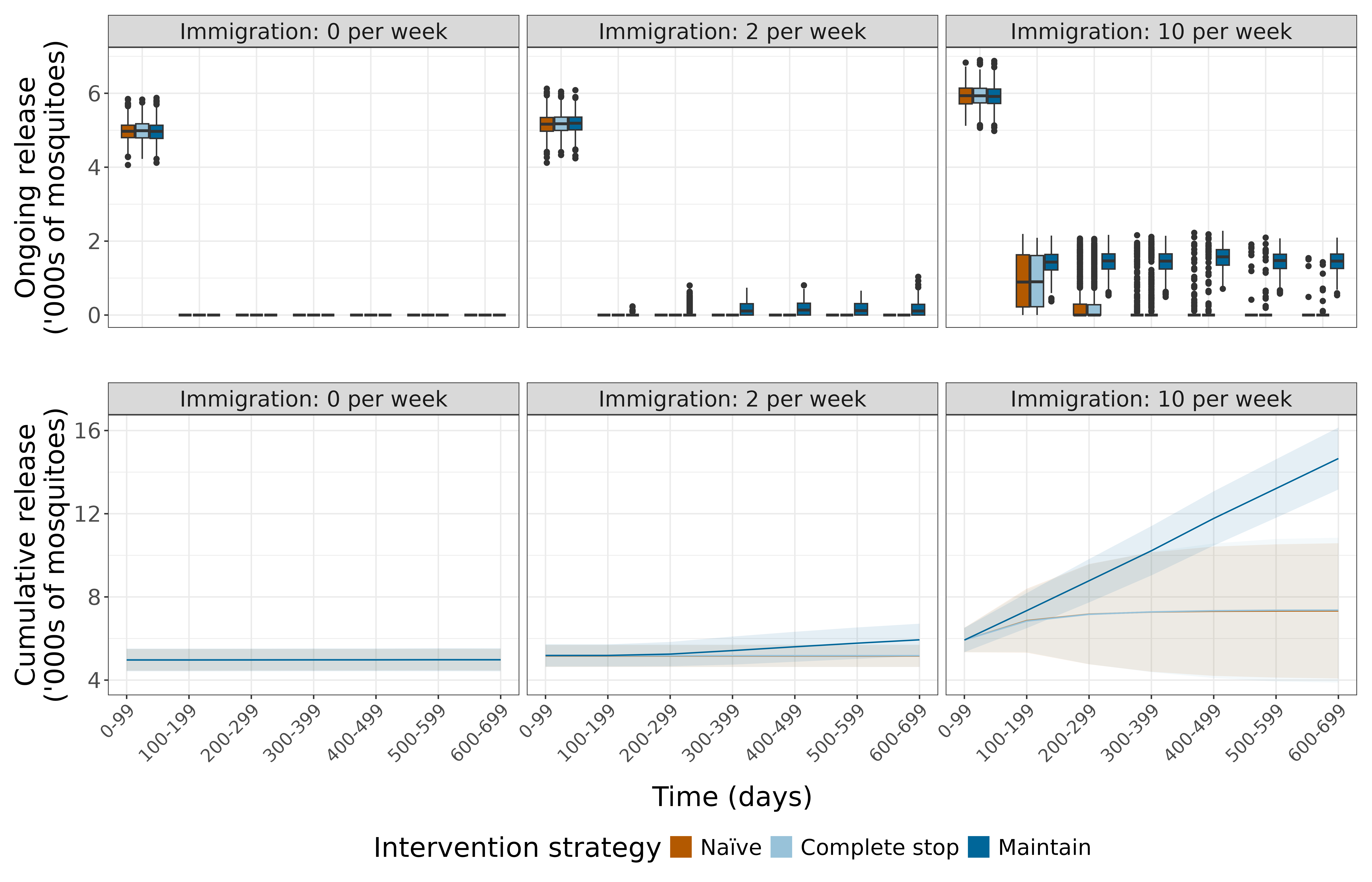}
    \caption{The release in thousands of \arwp\ mosquitoes as a proxy for cost under different intervention strategies and immigration rates. The $x$-axis represents the time in $100$-day intervals up until day $700$. The top row shows the ongoing cost every hundred days, measured as the number of mosquitoes released every $100$ days for each intervention strategy (na\"ive, complete stop, and maintain). The bottom row illustrates the cumulative release (in thousands of mosquitoes) over the entire period for each strategy. 
    The model parameters are the expected values defined in Supplementary Materials Table~\ref{supp-tab:parameters}.}
    \label{fig:intervention_costs}
\end{figure}

We compare the number of mosquitoes released for each scenario and immigration rate to compare different release scenarios (Figure \ref{fig:intervention_costs}). All release scenarios have a consistent cost in the first $100$ days of releases. Note that this is by design as all three scenarios run for at least $100$ days before considering any stopping conditions. The no immigration scenario acts as a comparison, with approximately $5,000$ adult mosquitoes released before ceasing at the $100$-day interval. Once the wild-mosquito population is suppressed in the low immigration scenario, it is only the maintain intervention strategy that continues to accumulate further adult releases, though these are in the hundreds of adults per week only and total approximately $6,000$ adults after $700$ days (Figure \ref{fig:intervention_costs}, middle column). When immigration is high, adult male \arwp\ releases need to match the wild immigration rate to maintain suppression on the \alb\ population. For the maintain intervention strategy, this overflooding rate increases linearly through time, accumulating to over $14,000$ adults over 700 days and is the most expensive of all strategies proposed.


Despite the increased ongoing cost of the maintain scenario, the results are more predictable.  In contrast, the na\"ive and complete stop scenarios have a higher variance. These results are qualitatively consistent for the low fitness parameters (Figure~\ref{supp-fig:intervention_costs_low}). For the high fitness parameters, the na\"ive scenario has the greatest ongoing and cumulative cost due to the increased difficulty in suppressing the wild \walbab\ \alb\ population (Figure~\ref{supp-fig:intervention_costs_high}). 
When age-related decay of CI is removed in the bi-directional model, there are no qualitative differences in the estimated cost of each scenario (Figure \ref{supp-fig:intervention_costs_nodecay}).

\section{Discussion}

As the Incompatible Insect Technique continues to be successfully deployed across the world, there is an increasing need to explore novel management strategies which suppress medically important mosquito populations. These strategies are broadening in scope to include complex landscapes and species with novel \textit{Wolbachia} strains and bi-directional CI effects. Traditionally, the release of females of an artificially-infected \textit{Wolbachia} strain through imperfect sex-separation would jeopardise the effectiveness of an IIT control program. However, when wild mosquito populations also exhibit a CI phenotype, we demonstrate, \textit{in silico}, how suppression and establishment of an artificial \textit{Wolbachia} strain is inhibited by immigration from a wild population.  

The Markov population process model extended here (from \cite{pagendam_modelling_2020}) allows for increasing complexity of both landscape and biological parameters, modelled stochastically around \textit{Wolbachia} establishment thresholds. Where previous authors have concerned themselves with questions around uncertainty of establishment through low levels of female contamination \cite{soh_strategies_2022, pagendam_modelling_2020, matsufuji_optimal_2023}, here we have intentionally explored the release of artificially-infected \textit{Wolbachia} female mosquitoes to understand the stability of establishment under a bi-directional CI system. Our approach also explored the inclusion of age-related decay of CI for adult male mosquitoes. Importantly, our results highlight the attributes of \textit{Wolbachia} IIT control programs that seek to maintain a wild \walba/\walbb\ \alb\ mosquito population under a high level of suppression while maintaining a manageable level of \arwp\ \textit{Wolbachia} establishment. These include: (i) higher \arwp\ invasion thresholds due to bi-directional CI impacting the efficiency of \textit{Wolbachia} establishment and (ii) release strategies that consider rates of wild mosquito population immigration into treatment blocks, with higher levels of immigration impacting both growth of the wild mosquito population and the efficiency of \arwp\ establishment. Attribute (i) highlights the flexibility a bi-directional CI system has on the success and failure of an IIT control program, while attribute (ii) starts to highlight the complex nature of biological systems, with immigration and emigration impacting outcomes of different release strategies.

We estimate a new unstable equilibrium threshold for establishment in a bi-directional, \arwp\ \textit{Wolbachia} system as approximately $40\%$ of the original population, smaller than the $50\%$ suggested by Moretti \textit{et al.} \cite{moretti_cytoplasmic_2018}.  Interestingly, the same conservative threshold for \arwp\ establishment is suggested when age-decay of CI is removed from the model. Our threshold is a conservative estimate, as only a small percentage ($0.7\%$) of simulations led to establishment at the $0.45$ initial proportion of \arwp\ (Figure~\ref{fig:cage_establish}). Our findings align with those observed in bi-directional CI IIT cage studies undertaken by Lombardi \textit{et al.} \cite{lombardi_incompatible_2024} which observed a slight decrease in the infection of \arwp\ infected females at approximately $40\%$ of the caged population. At close to double the establishment threshold of $22\%$ in an \textit{Ae. aegypti} uni-directional system \cite{axford_fitness_2016, pagendam_modelling_2020}, we suggest our estimated \alb\ threshold is primarily the result of inefficient mating (that is, adults of each \textit{Wolbachia} strain failing to find a compatible mate) in a well-mixed population. This observation is also based on differences in mosquito biology (e.g. fitness, average life expectancy, CI changes over time) between \textit{Ae. aegypti} and \alb. These \textit{in silico} simulations provided the establishment threshold for the intervention scenarios developed herein, which take into consideration immigration/emigration, female contamination rate and male release scenarios of an artificial \textit{Wolbachia} strain.  

The stable maintenance of population suppression is the primary goal of any mosquito IIT control strategy. 
Three management scenarios were chosen to represent different control decisions, primarily focused on whether \arwp\ releases continue or are ceased. Results demonstrate the logical end points of scenarios where: \arwp\ replacement occurs (no immigration, all three management scenarios); the wild-type population is allowed to recover (low and high immigration, na\"ive and complete stop scenarios); or ongoing management controlled both mosquito populations (low and high immigration, maintain scenarios). 

When we considered isolated populations with no immigration we observed low levels of \arwp\ establishment across all management scenarios.  This is consistent with previous modelling for \textit{Ae. aegypti} which suggests that even under high levels of suppression in an isolated population, establishment may be stochastically driven \cite{pagendam_modelling_2020}.
There was no qualitative difference between the complete stop and na\"{i}ve release strategies due to their shared stopping criteria, but both scenarios showed evidence of  reversibility of the IIT control program in the presence of immigration/emigration. Finally, the maintain scenario was the most effective at maintaining a stable wild-type population in the presence of immigration, reflecting a likely management scenario in the real world. While not simulated here, scaled interventions would likely cover larger urban areas, with multiple patches the size of our single population. Due to the strong effect of immigration rates on continual release and cost, these results suggest an \alb\ bi-directional CI intervention should aim to reduce immigration between patches to something more easily manageable across the entire area being treated, a result echoed in \textit{Ae. aegypti} management programs \cite{project_wolbachia_2021, montenegro2024wolbachia, crawford_efficient_2020}. 

IIT control programs need to consider the cost of male delivery to ensure the programs are competitive with traditional insecticide-based approaches. Here we model total mosquitoes as a proxy for cost, as the price per mosquito will vary depending on the costs related to rearing, sorting and labour to deliver individual male mosquitoes into a landscape. It is clear from these economic proxies that cost scales linearly for a small population with a carrying capacity of $420$ mosquitoes, and that it is relatively expensive to suppress a population that is constantly receiving a large weekly migration ($>2.5\%$ of the population) of individuals from outside the release area. Again, these results suggest that a low immigration rate between populations would reduce costs. We hypothesise that the effect of reduced immigration on cost is caused by the relative abundance of each \textit{Wolbachia} strain being approximately $50\%$ (Figure~\ref{fig:cage_establish}), causing a reduction in mating efficiency of adults finding compatible mates in a well-mixed population. However, the mating efficiency when considering competing \textit{Wolbachia} strains has not yet been explored in low abundance or densities.

Our results are robust to removing the age-related decay of CI in the \walba/\walbb\ population.  This suggests that the decay of CI of  \walba/\walbb\ mosquitoes does not substantially impact IIT control programs for these population sizes and modelled scenarios.  However, our \textit{in silico} cage experiments suggest that age-related CI decay may play a key role in mating efficiencies when the relative abundance of each \textit{Wolbachia} strain is $50\%$ (Figure \ref{fig:cage_establish}) or when considering larger populations (Figure \ref{supp-fig:cage_establish_high}).  Additionally, there are some nuanced qualitative differences in the control of \walba/\walbb\ populations over the two year period modelled here (Figure \ref{supp-fig:intervention_costs_nodecay}). Further research is needed to determine the effect of age-decay of CI on real world IIT control programs.  In particular, the interactions of age-related CI decay with seasonality, populations experiencing mass migration, or larger population sizes provide interesting areas for further study.

As a first conceptual model towards understanding these complex IIT releases, there are a number of limitations and opportunities for future work. We modelled simple release strategies every seven days, based on knowledge of the current mosquito population for an overflooding approach and a single modelled spatial location (or patch). However, the future success of the IIT relies on the critical need for the technology to scale efficiently and effectively over large areas. These findings should be modelled over a larger, meta-population scale typical of a large intervention.  Further, the interventions considered should be formally optimised to the desired outcomes of the program under cost and time requirements. When considering a meta-population approach, this optimisation will include identifying optimal spatial locations to manage release strategies, which has recently been considered for IIT control programs for \textit{Ae. aegypti} in Singapore \cite{lim_2024_wolbachia}.  If the objective is to reduce transmission, this will require a transmission model or consideration of vectorial capacity, a feature that is not considered here.  Finally, our suppression criteria of $10\%$ of the original population represents a conservative value based off a recommendation for a control program reducing nuisance biting in a North American context \cite{fonseca_area_2013}. Future estimates of this criteria should consider how high levels of suppression on a mosquito population impacts the vectorial capacity of a species and its ability to transmit pathogens between human hosts.

Robust planning and evidence-based science is essential when following the principle of ``\textit{Primum non nocere}'' or ``first, do no harm'' when trialling new technologies. One condition communities and regulators have insisted on for pioneering population suppression technologies such as \textit{Wolbachia} IIT and genetic engineering approaches is the ability to return a wild insect population to an original state. 
The release of females through imperfect sex separation technologies has been considered an unwanted outcome of the IIT in the past. However, we suggest that immigration of a wild \textit{Wolbachia} strain (\walba/\walbb) can be used to suppress the establishment of an artificially introduced \textit{Wolbachia} strain (\arwp). Further, we have demonstrated that the introduced strain can be eradicated if immigrants (or inflows of the original strain) are utilised to return the targeted insect population to a previous natural state. This was a common outcome in all management scenarios where immigration was present and \textit{Wolbachia} (\arwp) releases had ceased; the higher the immigration rate, the quicker the population approached its original natural steady state. Returning an insect population to a previous natural state is an outcome that may be desirable by local communities opting out of a \textit{Wolbachia} program, or if a commercial program is no longer financially viable.

\section*{CRediT statement}
MR contributed to Methodology, Software, Validation, Formal analysis, Investigation, Writing - Original Draft, Writing - Review \& Editing, and Visualisation. MM contributed to Software, Validation, Formal analysis, Investigation, Writing - Original Draft, Writing - Review \& Editing, and Visualisation. DP contributed to Conceptualisation, Methodology, Software, and Writing - Review \& Editing. RH contributed to Conceptualisation, Methodology, Software, Validation, Formal analysis, Writing - Original Draft, Writing - Review \& Editing, and Visualisation. BT contributed to Conceptualisation, Methodology, Validation, Formal analysis, Writing - Original Draft, Writing - Review \& Editing, Visualisation, and Funding acquisition.

\section*{Acknowledgements}

We would like to acknowledge Maurizio Calvitti and Riccardo Moretti for their input in early discussions on the bidirectional CI system conceptualised here.

\section*{Funding}

This paper was partially funded by a National Health and Medical Research Council grant 20122404.

\section*{Conflicts of interest}

The Authors declare that there is no conflict of interest.

\section*{Data and code availability}

The code is available at \url{https://github.com/Matthew-Ryan1995/bidirectional_IIT_control_model}.  The only data used was parameter values, available from both Table S1 and the code repo.

\bibliographystyle{vancouver}
\bibliography{refs}

\end{document}


\maketitle

\section{Model details}

Here the full model details are described, with Table \ref{tab:parameters} outlining the model parameters and values, and Table \ref{tab:transitions} providing the full model transitions for the continuous time Markov chain (CTMC) implementation of our Markov population process. When referring to ``high fitness parameters'' the upper bounds for the number of immature classes, the proportion of mated females and the carrying capacity were used, and the lower bounds for the birth and death rates were used (Table \ref{tab:parameters}, note the markers).  This choice for high fitness parameters means the mosquitoes live longer lives in larger populations. For ``low fitness parameters'' the other extrema were used, such that the mosquito lives were on average shorter and populations smaller.

\begin{centering}
\begin{longtable}{c|>{\raggedright\arraybackslash}p{5.2cm} |>{\raggedright\arraybackslash}p{3.3cm} | l}
    \caption{Descriptions and values of model parameters. All rates are in units of per day unless otherwise specified. $\dagger$ high fitness parameters. $\flat$ low fitness parameters.  CI - Cytoplasmic incompatibility.} 
    \label{tab:parameters} \\ \hline 
    Symbol & Description & Value (min, max) & Source \\ \hline 
    \endfirsthead
    \multicolumn{4}{c}
    {\tablename\ \thetable\ -- \textit{Continued from previous page}} \\
    \hline
    Symbol & Description & Value (min, max) & Source \\
    \hline
    \endhead
    \hline \multicolumn{4}{r}{\textit{Continued on next page}} \\
    \endfoot
    \hline
    \endlastfoot
    $K$ & Number of male classes; average number of days male CI changes & 20 & \cite{calvitti_wolbachia_2015}\\
    $\sigma$ & CI changing rate & $1/K$ & \\
    $k$ & Number of immature classes; average number of days for an egg to mature into an adult & 12 (10$^\flat$, 50$^\dagger$) & \cite{alto_precipitation_2001} \\
    $\gamma$ & Maturation rate of immature classes & $1/k$&  \\
    $p_\text{mated}$ & Proportion of females mated at steady state & 0.800 (0.500$^\flat$, 0.800$^\dagger$) &  \\
    $C$ & Total adult population at steady state & 420 (400$^\flat$, 800$^\dagger$) & \cite{manica_spatial_2016} \\
    $\lambda$ & Intrinsic birth rate of females & 0.253 (0.106$^\dagger$, 0.400$^\flat$) & \cite{alto_temperature_2001, nur_aida_population_2008} \\
    $p_f$ & Proportion of immatures that become adult females & 0.500 &  \\
    $p_m$ & Proportion of immatures that become adult males & $1-p_f$ & \\
    $\mu_M$ & Death rate of adult males & 0.128 (0.043$^\dagger$, 0.230$^\flat$) & \cite{vavassori_active_2019} \\
    $\mu_F$ & Death rate of adult females & 0.116 (0.006$^\dagger$, 0.200$^\flat$) & \cite{vavassori_active_2019} \\
    $\zeta^F=\zeta^M$ & Overall male/female immigration rate & 0.286 (0, 1.429) &  \\
    $\xi^M$, $\xi^F$ & Emigration rates & Eq \eqref{eqn:im_em_eq} & \\
    $I_\text{max}$ & Carrying capacity of immatures & Eq \eqref{eqn:Imax} & \cite{manica_spatial_2016} \\
    $\eta$ & Mating rate between adults & Eq \eqref{eqn:eta} & \\
    $c_{\walbab\times \arwp, l}$ & CI leading to proportion of viable offspring by mated \walbab\ females with \arwp\ males of age $l$ & 1 $\forall l$ & \cite{calvitti_wolbachia_2015} \\
    $c_{\arwp\times \walbab, l}$ & CI leading to proportion of viable offspring by mated \arwp\ females with \walbab\ males of age $l$ & 1 ($l=1$--$14$), 0.68 ($l=15$--$19$), 0 ($l\geq20$) & \cite{calvitti_wolbachia_2015} \\
    $\text{Fried}_{\walbab}, \text{Fried}_{\arwp}$ & Fried's index for \arwp\ and \walbab. & 1 &\\
    \bottomrule
\end{longtable}
\end{centering}

\begin{table}[htbp]
    \caption{Stoichiometries of the full model.}
    \label{tab:transitions}
    \centering
    \begin{tabular}{l|l | r}
    \toprule
    Event & State change & Transition rates\\
    \midrule
    Birth of an immature & $I_{w, 1} + 1$ & $\tilde{\lambda} \bar{F}_{w}$ \\
    Ageing of immatures & $(I_{w, i}, I_{w, i+1}) + (-1, 1)$ &  $k \gamma I_{w, i}, \forall i\in[1,k-1]$ \\
    Maturation into male & $(I_{w, k}, M_{w, 1}) + (-1, 1)$ & $p_m k \gamma I_{w, k}$ \\
    Maturation into female & $(I_{w, k}, F_{w}) + (-1, 1)$ & $p_f k \gamma I_{w, k}$\\
    Ageing of male & $(M_{w, i}, M_{w, i+1}) + (-1, 1)$ & $K\sigma M_{w, i}$\\
    Death of male & $M_{w, i} - 1$ & $\mu_M M_{w, i}$\\
    Death of unmated female & $F_{w} - 1$ & $\mu_F F_{w}$ \\
    Mating & $(F_{w}, F_{w\times v, l}) + (-1, 1) $ & $\eta F_{w} M_{v, l}$\\
    Death of mated female & $F_{w\times v, l} - 1$ & $\mu_F F_{w\times v, l}$\\
    Emigration of male & $M_{w, i}-1$ & $\xi_i^M M_{w, i}$ \\
    Emigration of unmated female & $F_w-1$ & $\xi_0^F F_{w}$\\
    Emigration of mated female & $F_{w\times v, l}-1$ & $\xi_l^F F_{w\times v, l}$\\
    Immigration of male & $M_{w, i}+1$ & $\zeta_i^M $ \\
    Immigration of unmated female & $F_w+1$ & $\zeta_0^F $\\
    Immigration of mated female & $F_{w\times v, l}+1$ & $\zeta_l^F $\\
    \bottomrule
    \end{tabular}
\end{table}

\section{Model generalisation}

\begin{figure}[htbp]
    \centering
    \includegraphics[width=\linewidth]{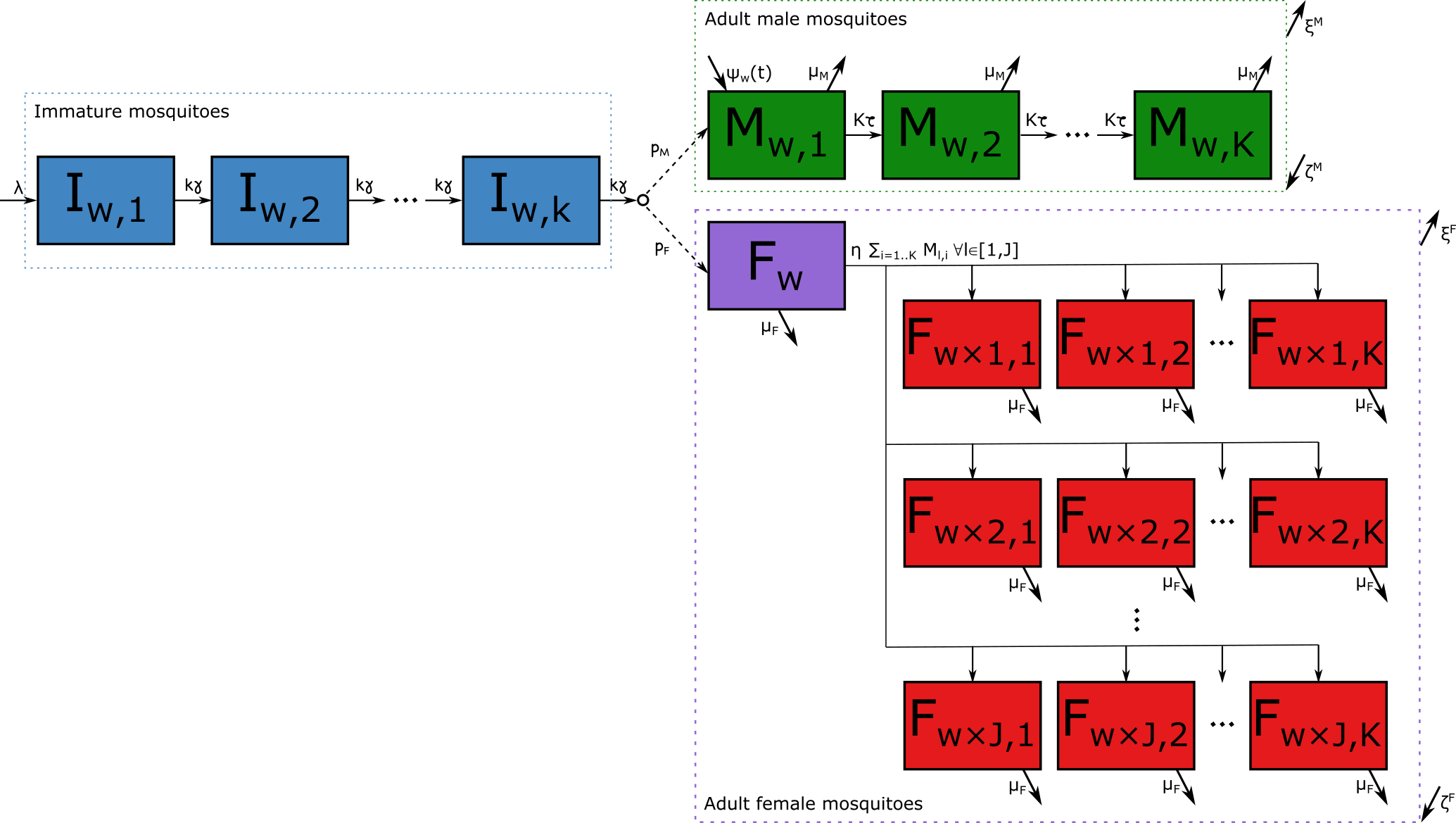}
    \caption{Schematic of the generalised mosquito population model for $J$ strains of \textit{Wolbachia}, where this depicts the progression for a single strain $w$ of the Immature stages (blue) of strain $w$ through to different male ages (green) or females who are unmated (purple) then mated with males with a specific \textit{Wolbachia} strain and age (red). This is then repeated for the other $J-1$ strains. Immigration and emigration occurs from each adult compartment, captured here by the flows from the dashed boxes grouping compartments.}
    \label{fig:general_compartmental_diagram}
\end{figure}

A generalised version for $J$ strains of \textit{Wolbachia} is depicted in Figure~\ref{fig:general_compartmental_diagram}. The population progression is from ``immature'' ($I_{w,i} \forall w\in[1,\ldots,J]$ and $i\in[1,k]$) to adult male ($M_{w,j}$) or adult females that are unmated ($F_{w}$) or mated with males with strain $v$ when the males were age $l$ ($F_{w\times v, l}$).

\section{Wild-type steady states}
\label{sec:wild_ss}

To explore the intervention scenarios, we assume that the wild-type mosquito (\walbab) population is at an equilibrium before the introduction of any new species.  Here, as done by \pag, we determine the equilibrium states for the wild-type species with no invasive species added based on assumed observable mosquito population values. In \S\ref{sec:closed} we assume a closed population model with no immigration or emigration, which is then added in \S\ref{sec:imm}. We determine biologically feasible parameter constraints in \S\ref{sec:constraints}.

\subsection{Closed population steady states}\label{sec:closed}

We apply an ordinary differential equation (ODE) approximation of the CTMC since we are studying a steady state. The following ODEs describe the dynamics, where dependence on \textit{Wolbachia} strain as been dropped for simplicity of notation and $\dot{X}$ denotes the derivative of compartment $X$ with respect to time:

\begin{align}
    \label{eqn:1}
    \dot{I}_{1} &= \tilde{\lambda} \bar{F}_{M} - k \gamma I_{1} \\
    \label{eqn:2}
    \dot{I}_{i} & = k \gamma I_{i - 1} - k \gamma I_{i}\,,\, i = 2, 3, \dots, k\\
    \label{eqn:3}
    \dot{M}_{1} & =  k \gamma \prm I_{k} - (\mu_M + K \sigma) M_1 \\
    \label{eqn:4}
    \dot{M}_{j} & = K \sigma M_{j-1} - (\mu_M + K \sigma) M_j\,,\, j = 2, 3, \dots, K-1 \\
    \label{eqn:5}
    \dot{M}_{K} & = K \sigma M_{k-1} - \mu_M M_K \\
    \label{eqn:6}
    \dot{F} & = k \gamma \prf I_{k} - (\mu_F + \tilde{\eta}) F \\
    \label{eqn:7}
    \dot{F}_{M, l} &= \eta M_{l} F - \mu_F F_{M, l}\, ,\, l = 1, 2, \dots, K,
\end{align}
where
\begin{itemize}
    \item $F_{M, l}$ is the mated female class with males of class $l$, 
    \item $\tilde{\eta}$ is the basic mating rate given by $\tilde{\eta} = \eta \bar{M}$, $\bar{M} = \sum\limits_{j=1}^K M_j$, and
    \item $\tilde{\lambda}$ is the density-dependant birth rate given by
    \[
        \tilde{\lambda} = \lambda\frac{\left(I_{max} - I_{total}\right)}{I_{max}}\, ,
    \]
    with $I_{max}$ being the carrying capacity of larvae and $I_{total} = \sum\limits_{i = 1}^k \left(  I_{i} \right)$.
    \item $\sigma$ relates to age-related loss of cytoplasmic incompatibility by male mosquitoes, designed such that $K \sigma=1$ for our scenarios.
    \item $\prf$ and $\prm$ are the probabilities an immature mosquito matures into an adult male or female, which must add to one. For the strains we study, $\prf=\prm=1/2$.
\end{itemize}

Suppose we are given the maturation rate $\gamma$, birth rate $\lambda$, death rates $\mu_M$ and $\mu_F$, the proportion of mated female mosquitoes at equilibrium $p_{mated}$, and equilibrium capacity of the adult population $C$.  Then, following \pag, we can calculate all equilibrium states and unknown parameters of the System \eqref{eqn:1}--\eqref{eqn:7}.  In the usual way, setting the rates of change (left hand sides) of Equations \eqref{eqn:1} and \eqref{eqn:2} to zero gives steady state solutions denoted with a hat:
\[
     k \gamma \hat{I}_{1} = \tilde{\lambda} \bar{F}_M
\]
and
\[
    \hat{I}_{i} = \hat{I}_{1}
\]
for all $i = 1, 2, \dots, k$.  This implies that
\[
    \hat{I}_{i} = \bar{I}
\]
for a constant $\bar{I}$ independent of $i$.  Equations \eqref{eqn:3}--\eqref{eqn:5} give
\begin{align*}
    \bar{I} &= \frac{\mu_M + K \sigma}{k \gamma \prm} \hat{M}_1\, , \\
    \hat{M}_{j} & = \frac{K \sigma}{(\mu_M + K \sigma)} \hat{M}_{j-1}\, ,\text{ for } j = 2, 3, \dots, K-1,\text{ and,}\\
    \hat{M}_{K} & = \frac{K \sigma}{\mu_M} \hat{M}_{K-1}\, .
\end{align*}
Writing $\varphi = \frac{K \sigma}{\mu_M + K \sigma}$ gives
\begin{align}
    \bar{I} &= \frac{K \sigma}{\varphi k \gamma \prm} \hat{M}_1\, , \label{eq:Ibar} \\
    \hat{M}_{j} & = \varphi^{j-1} \hat{M}_{1}\, ,\text{ for } j = 1, 2, 3, \dots, K-1,\text{ and,}\\
    \hat{M}_{K} & = \frac{K \sigma \varphi^{K-2}}{\mu_M} \hat{M}_{1}\, .
\end{align}
Hence, once we know $\hat{M}_1$ we know all immature states and all male states.  To calculate $\hat{M}_1$, we first consider the total adult population at equilibrium
\begin{equation}
    \label{eqn:total_capacity}
    \bar{M} + \hat{F} + \bar{F}_M = C\, ,
\end{equation}
where
\[
    \bar{M} = \sum_{j=1}^K \hat{M}_j
\]
and
\[
    \bar{F}_M = \sum\limits_{l = 1}^K \hat{F}_{M, l}\, .
\]
Noting that, at equilibrium, for the adult populations to remain constant the total deaths must equal the new maturations, such that
\begin{align*}
    \mu_M \bar{M} & = k \gamma \prm \hat{I}_k\,, \text{ and}\\
    \mu_F\left( \hat{F} + \bar{F}_M \right) & = k \gamma \prf \hat{I}_k\, ,
\end{align*}
which means
\[
    \frac{\mu_M \bar{M}}{\prm} = \frac{\mu_F(\hat{F} + \bar{F}_M)}{\prf}\, .
\]
This gives that the ratio
\begin{equation}
    \label{eqn:theta}
    \frac{\hat{F} + \bar{F}_M}{\bar{M}} = \frac{\mu_M \prf}{\mu_F \prm} = \theta\, .
\end{equation}
Combining Equations \eqref{eqn:total_capacity} and \eqref{eqn:theta} gives that
\begin{align}
    \label{eqn:Mbar}
    \bar{M} & = \frac{C}{1+ \theta}\,, \\
    \label{eqn:Fbar}
    \hat{F} & = \frac{\theta}{1+\theta}(1-p_{mated}) C\, , \text{ and}\\
    \label{eqn:Fbar_mated}
    \bar{F}_M &= \frac{\theta}{1+\theta}p_{mated} C\, ,
\end{align}
where $p_{mated}=\bar{F}_M/(\hat{F}+\bar{F}_M)$. Recalling that $\bar{M} = \sum\limits_{j = 1}^K \hat{M}_i$, then 
\begin{align*}
    \frac{C}{1+ \theta} & = \bar{M}\\
        & = \left[\sum_{j = 1}^{K-1} \varphi^{j-1} + \frac{K \sigma \varphi^{K-2}}{\mu_M}\right]  \hat{M}_{1} \\
        & = \left[\frac{1 - \varphi^{K-1}}{1 - \varphi} + \frac{K \sigma \varphi^{K-2}}{\mu_M}\right] \hat{M}_{1}\, .
\end{align*}
Hence
\begin{equation}
    \label{eqn:M1}
    \hat{M}_{1} = \frac{C}{\left(1+ \theta\right)\left[\frac{1 - \varphi^{K-1}}{1 - \varphi} + \frac{K \sigma \varphi^{K-2}}{\mu_M}\right]}\, .
\end{equation}
Now only two quantities remain unknown: the larvae population limits $I_{max}$ and the mating rate $\eta$.  To find $I_{max}$, we equate Equation \eqref{eqn:1} to zero giving
\begin{align*}
    \bar{I} & = \frac{\tilde{\lambda}}{k \gamma} \hat{F}_M \\
    & = \frac{\hat{F}_M \lambda}{k \gamma}\frac{\left(I_{max} - \hat{I}_{total}\right)}{I_{max}}\, ,
\end{align*}
where $\hat{I}_{total} = \sum\limits_{i = 1}^k \hat{I}_{i} = k \bar{I}$.  Rearranging for $I_{max}$ then gives
\begin{align}
    \label{eqn:Imax}
    I_{max} & = \dfrac{k\bar{I}}{1 - \dfrac{k \gamma \bar{I}}{\lambda \bar{F}_M}}\, .
\end{align}
Finally, to calculate the mating rate $\eta$ we set Equation \eqref{eqn:6} to zero to get
\[
    k \gamma \prf \bar{I} = \mu_F \hat{F} + \eta \bar{M} \hat{F}
\]
and so
\begin{equation}
    \label{eqn:eta}
    \eta = \frac{k \gamma \prf \bar{I} - \mu_F \hat{F}}{\bar{M} \hat{F}}\, .
\end{equation}
From Equation \eqref{eqn:7},
\begin{equation}
    \hat{F}_{M, l} = \frac{\eta \hat{M}_{l} \hat{F}}{\mu_F}\, ,
\end{equation}
and so all steady state values can be determined.

\subsection{Immigration and emigration}\label{sec:imm}

We introduce immigration and emigration into the system such that the steady states found in Section \ref{sec:wild_ss} remain unchanged.  Assuming no immigration at the immature stages and denoting immigration by $\zeta^s$ and emigration by $\xi^s$ ($s \in \{M, F\}$), Equations \eqref{eqn:1}-\eqref{eqn:7} become:

\begin{align}
    \label{eqn:1a}
    \dot{I}_{1} &= \tilde{\lambda} \bar{F}_{M} - k \gamma I_{1} \\
    \label{eqn:2a}
    \dot{I}_{i} & = k \gamma I_{i - 1} - k \gamma I_{i}\,,\, i = 2, 3, \dots, k\\
    \label{eqn:3a}
    \dot{M}_{1} & =  k \gamma \prm I_{m, k} - (\mu_M + K \sigma) M_1 + \zeta^M_1 - \xi^M_1 M_1 \\
    \label{eqn:4a}
    \dot{M}_{j} & = K \sigma M_{j-1} - (\mu_M + K \sigma) M_j + \zeta^M_j - \xi^M_j M_j \,,\, j = 2, 3, \dots, K-1 \\
    \label{eqn:5a}
    \dot{M}_{K} & = K \sigma M_{k-1} - \mu_M M_K + \zeta^M_K - \xi^M_K M_K \\
    \label{eqn:6a}
    \dot{F} & = k \gamma \prf I_{f, k} - (\mu_F + \tilde{\eta}) F +\zeta_0^{F} - \xi_0^{F} F\\
    \label{eqn:7a}
    \dot{F}_{M, l} &= \eta M_{l} F - \mu_F F_{M, l} +\zeta_{l}^{F} - \xi_l^{F} F_{M, l} \, ,\, l = 1, 2, \dots, K,
\end{align}
To maintain the steady states of Section \ref{sec:wild_ss}, we require at equilibrium that
\begin{align}
    \label{eqn:im_em_eq}
    \zeta_j^M & = \xi_j^M \hat{M}_j\, , j = 1, 2, \dots, K, \text{ and}\\
    \zeta_l^F & = \xi_l^M \hat{F}_{M, l}\, , l = 0, 1, 2, \dots, K,
\end{align}
where $\hat{F}_{M, 0} = \hat{F}$.  We first focus on the immigration rate. We assume the following:
\begin{enumerate}
    \item[A1] There are overall immigration per capita rates for males and females.
    \item[A2] Immigration is from an external patch with equal population size which is already in an equilibrium state.
    \item[A3] The rate of immigration depends on mosquito age, not mate status.
\end{enumerate}
Looking at the male immigration rate, Assumption A1 tells us that there is a constant $\zeta^M$ such that
\[
    \sum_{j = 1}^K \zeta_j^M = \zeta^M\, ,
\]
whereas Assumption A2 says
\[
    \zeta_j^M = \tilde{\zeta}_j^M \hat{M}_j\, ,
\]
that is, $\zeta_j^M$ is proportional to the number of male mosquitoes aged $j$ in the steady state solutions.  Combining these gives
\begin{align*}
    \zeta^M & =  \sum_{j = 1}^K \zeta_j^M \\
        & = \sum_{j = 1}^K \tilde{\zeta}_j^M \hat{M}_j \\
        & = \left(\sum_{j = 1}^{K-1} \left(\tilde{\zeta}_j^M \varphi^{j-1}\right) + \dfrac{K \sigma \varphi^{K-2}}{\mu_M}\tilde{\zeta}_{K}^M\right)\hat{M}_1\, ,
\end{align*}
which gives
\begin{equation}
    \label{eqn:imm_M_1}
    \left(\sum_{j = 1}^{K-1} \left(\tilde{\zeta}_j^M \varphi^{j-1}\right) + \dfrac{K \sigma \varphi^{K-2}}{\mu_M}\tilde{\zeta}_{K}^M\right) = \frac{\zeta^M}{\hat{M}_1}\, .
\end{equation}
Now, even with the constraint that $\tilde{\zeta}_j^M \geq 0$ for all $j = 1, 2, \dots, K$, Equation \eqref{eqn:imm_M_1} defines a whole surface of solutions.  Thus, we constrain the problem with Assumption A3 and assume that
\begin{equation}
    \label{eqn:simplify_zeta}
    \tilde{\zeta}_j^M = \tilde{\zeta}^M P_j\, ,
\end{equation}
where $\tilde{\zeta}^M$ is constant and $P_j$ is the probability that a male mosquito lives to be age $j$.  Let $Y$ denote the time a male mosquito is alive, which from adding Equations \eqref{eqn:3}--\eqref{eqn:5} we find is exponentially distributed with rate $\mu_M$.  Then
\begin{align*}
    P_j & = P(m \in M_j) \\
    & = P\left(\frac{j - 1}{K\sigma} < Y \leq \frac{j}{K\sigma}\right) \\
    & = e^{-\mu_M (j -1) / (K\sigma)} - e^{-\mu_M j / (K\sigma)} \, .
\end{align*}
Note we have divided the index $j$ by the transition between male age classes $K\sigma$. Since $j \leq K$, when $j=K$ we get
\[
    P_K = P\left(Y > \frac{j-1}{ K\sigma}\right)\, .
\]
Incorporating Equation \eqref{eqn:simplify_zeta} into \eqref{eqn:imm_M_1} gives that
\begin{equation}
    \tilde{\zeta}^M = \dfrac{\zeta^M}{\hat{M}_1\left(\sum_{j = 1}^{K-1} \left(P_j \varphi^{j-1}\right) + \dfrac{K \sigma \varphi^{K-2}}{\mu_M}P_K\right) }\, .
\end{equation}
Using Equation \eqref{eqn:im_em_eq} gives that the emigration rate is
\[
    \xi_j^M = \tilde{\zeta}_j^M
\]
for all $j = 1, 2, \dots, K$.  Now consider the female immigration rate.  Again, Assumptions A1 and A2 give us that
\[
    \sum_{l = 0}^K \tilde{\zeta}_l^F \hat{F}_{M, l} = \zeta^F\, ,
\]
but now Assumption A3 tells us that this only depends on mosquito age and not mating status.  Since we only consider one age strata for female mosquitoes, Assumption A3 implies that there is a constant $\tilde{\zeta}^F$ such that
\[
    \tilde{\zeta}_l^F = \tilde{\zeta}^F
\]
for all $l = 0, 1, 2, \dots, K$.  Thus, 
\[
    \tilde{\zeta}^F = \dfrac{\zeta^F}{ \sum_{l = 0}^K \hat{F}_{M, l}}\, .
\]
Noting that
\begin{align*}
    \sum_{l = 0}^K \hat{F}_{M, l} & = \hat{F} + \bar{F}_M \\
        & = \frac{\theta}{1 + \theta} C\, ,
\end{align*}
we get
\begin{equation}
    \label{eqn:immigration_female}
    \tilde{\zeta}^F = \frac{(1 + \theta) \zeta^F}{\theta C}\, .
\end{equation}
Again, Equation \eqref{eqn:im_em_eq} gives that the emigration rate is
\[
    \xi_l^F =  \tilde{\zeta}^F
\]
for all $l = 0, 1, 2, \dots, K$. In summary and for clarity, we get that
\begin{align*}
    \zeta_j^M & =  \dfrac{\zeta^M P_j \hat{M}_j}{\hat{M}_1\left(\sum_{i = 1}^{K-1} \left(P_i \varphi^{i-1}\right) + \dfrac{K \sigma \varphi^{K-2}}{\mu_M}P_K\right) } \, ,\\
    \xi_j^M & = \dfrac{\zeta^M P_j}{\hat{M}_1\left(\sum_{i = 1}^{K-1} \left(P_i \varphi^{i-1}\right) + \dfrac{K \sigma \varphi^{K-2}}{\mu_M}P_K\right) } \, ,\\
    \zeta_l^F & =  \frac{(1 + \theta) \zeta^F}{\theta C} \hat{F}_{M, l}, \\
    \xi_l^F & = \frac{(1 + \theta) \zeta^F}{\theta C} \, ,
\end{align*}
for $j = 1, 2, \dots, K$ and $l = 0, 1, 2, \dots, K$.

\subsection{Biologically feasible constraints on parameters}\label{sec:constraints}

The calculation of the secondary parameters in Section \ref{sec:wild_ss} gives a constraint on the parameters for the model to give biologically feasible results. Equation \eqref{eqn:Imax} gives that
\[
    \frac{k \gamma \bar{I}}{\lambda \bar{F}_M} < 1\, .
\]
First considering the numerator, substituting $\bar{I}$ from \eqref{eq:Ibar} gives 
\begin{align*}
    k \gamma \bar{I} & = k \gamma \frac{K \sigma }{\varphi k\gamma \prm} \hat{M}_1\\
        & = \frac{K\sigma C}{\prm \left(1+ \theta\right) \varphi \left[\frac{1 - \varphi^{K-1}}{1 - \varphi} + \frac{K \sigma \varphi^{K-2}}{\mu_M}\right]}\, .
\end{align*}
Noting that $1-\varphi = (\mu_M \varphi)/(K \sigma)$, this reduces to
\begin{equation}
    \label{eqn:gamma_Ibar}
   k \gamma \bar{I} = \frac{C \mu_M}{\prm (1+\theta)}\, .
\end{equation}
Considering the denominator, substitute in Equation~\eqref{eqn:Fbar_mated},
\begin{align*}
   \lambda \bar{F}_M & = \lambda p_{mated} C \frac{\theta}{1+\theta}\, .
\end{align*}
Hence,
\begin{align*}
    \frac{k \gamma \bar{I}}{\lambda \bar{F}_M} & = \dfrac{\frac{C \mu_M}{\prm(1+\theta)}}{\lambda p_{mated} C \frac{\theta}{1+\theta}} \\
        & = \frac{\mu_M}{\prm \lambda p_{mated} \theta}.
\end{align*}
Recalling that $\theta = \frac{\mu_M \prf}{\mu_F \prm}$ gives the constraint
\begin{equation}
    \label{eqn:constraint_1}
    \frac{\mu_F}{\prf \lambda p_{mated}} < 1\, .
\end{equation}
This effectively just states that female mosquitoes must be born (noting immatures all become adults) at a greater rate than they die. The strict inequality is likely a consequence of the density-dependence (i.e. due to $\tilde{\lambda}$). 

Note that there seems to be a second constraint from Equation \eqref{eqn:eta}, which requires that
\[
    \frac{k \gamma \prf \bar{I}}{\mu_F \hat{F}} > 1\, .
\]
However, combining Equations \eqref{eqn:gamma_Ibar}, \eqref{eqn:Fbar}, and \eqref{eqn:theta} gives that
\[
    \frac{k \gamma \prf \bar{I}}{\mu_F \hat{F}} = \frac{1}{1-p_{mated}}\, .
\]
Hence, this constraint is met for all biologically feasible $p_{mated} \in (0, 1)$.

\subsection{Multiple strains of \textit{Wolbachia}}

When considering $J\geq 2$ strains of \textit{Wolbachia} we modify the stoichiometries of Table \ref{tab:transitions} in the following ways.  First, all immature mosquitoes, regardless of \textit{Wolbachia} strain, contribute to the total immature pool $I_{total}$.  This feature creates a resource competition between the multiple strains of \textit{Wolbachia}.  Second, the mating rate $\eta$ (Equation \ref{eqn:eta}) between a female of strain $w$ and a male of age $l$ strain $v$ is modified by  Fried's index and mating competitiveness.  Mating pressure for type $v$ and age $l$ is defined as
\[
    \kappa_{v, l} = \text{Fried}_{v} M_{v, l}.
\]
Then the mating rate is modified by the mating competitiveness between age and strain as
\[
    \eta_{v, l} = \eta \frac{\kappa_{v, l}}{\sum_{v, l} \kappa_{v, l}}\, .
\]
This modification means that mosquito mating competitiveness is proportional to their density in the population.  The final modification is through the birthing rate.  When mosquitoes of two different strains $w$ and $v$ of \textit{Wolbachia} mate there is an effect of CI that reduces offspring viability.  This is represented in our model as modifying the birth rate $\tilde{\lambda}$ as 
\[
    \tilde{\lambda}_{w \times v, l} = (1-c_{w \times v, l}) \tilde{\lambda}\, ,
\]
where $c_{w \times v, l}$ is the CI between \textit{Wolbachia} strains $w$ and $v$ for an adult male of age $l$.

\section{Additional figures and tables}


\begin{figure}[H]
    \centering
    \includegraphics[width=\linewidth]{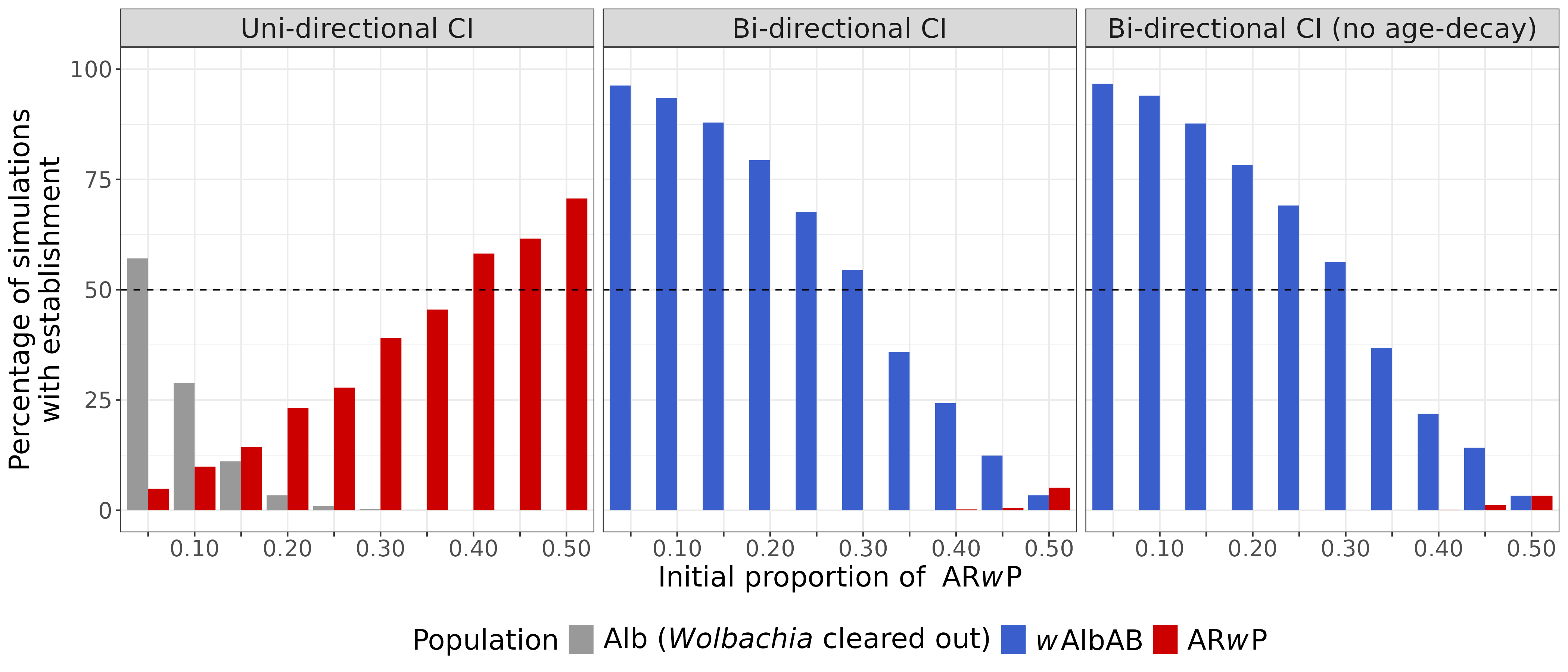}
    \caption{Percentage of simulations with establishment of each population at the end of the $500$ \textit{in silico} cage simulations. A population is established if the proportion of adults is greater than $10\%$ of the initial population size. The left panel shows simulations with uni-directional Cytoplasmic Incompatibility (CI), the middle panel depicts bi-directional CI with age-decay and the right panel bi-directional CI without age-decay. Red columns represent the \arwp\ population, blue columns the \walbab\ population, and grey columns the \textit{Wolbachia} cleared out wild-type population. The $x$-axis indicates the initial proportion of \arwp, and the $y$-axis shows the percentage of simulations that resulted in establishment. The dotted horizontal line marks $50\%$ of simulations. Each simulation started with $420$ adult mosquitoes (half male, half female), and the fitness parameters are the low values defined in Table~\ref{tab:parameters}.}
    \label{fig:cage_establish_low}
\end{figure}


\begin{figure}[H]
    \centering
    \includegraphics[width=\linewidth]{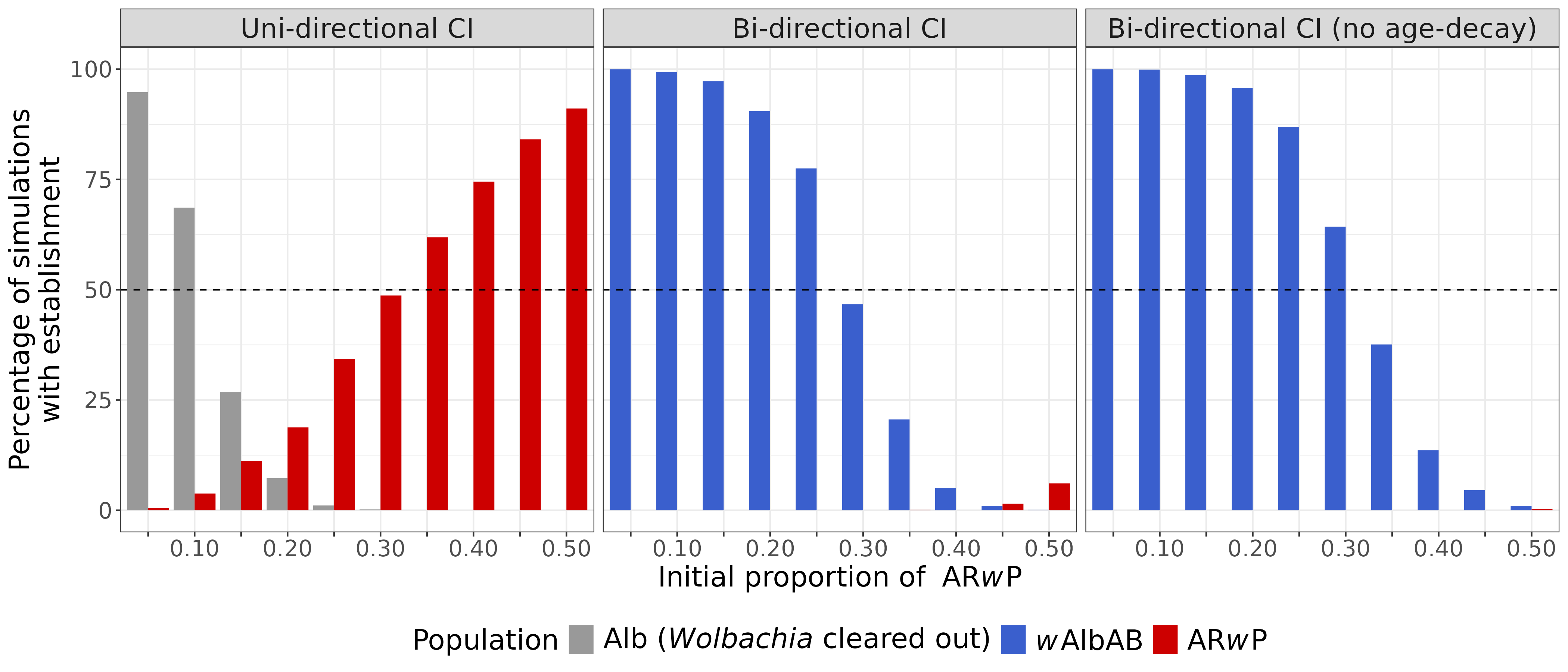}
    \caption{Percentage of simulations with establishment of each population at the end of the $500$ \textit{in silico} cage simulations. A population is established if the proportion of adults is greater than $10\%$ of the initial population size. The left panel shows simulations with uni-directional Cytoplasmic Incompatibility (CI), the middle panel depicts bi-directional CI with age-decay and the right panel bi-directional CI without age-decay. Red columns represent the \arwp\ population, blue columns the \walbab\ population, and grey columns the \textit{Wolbachia} cleared out wild-type population. The $x$-axis indicates the initial proportion of \arwp, and the $y$-axis shows the percentage of simulations that resulted in establishment. The dotted horizontal line marks $50\%$ of simulations. Each simulation started with $420$ adult mosquitoes (half male, half female), and the fitness parameters are the high values defined in Table~\ref{tab:parameters}.}
    \label{fig:cage_establish_high}
\end{figure}

\begin{figure}[H]
    \centering
    \includegraphics[width=\linewidth]{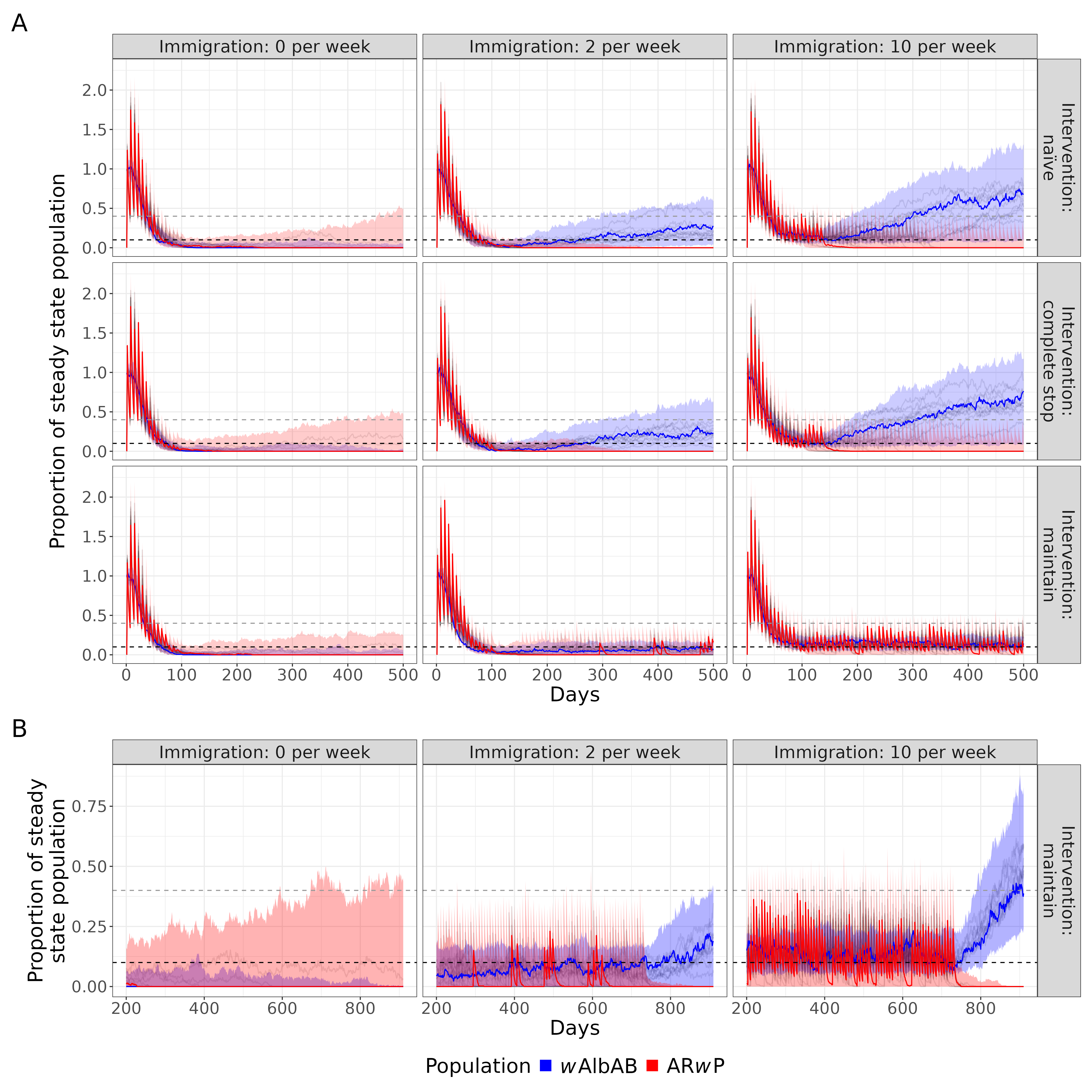}
    \caption{Proportion of mosquitoes relative to the steady-state population over time for different release strategies (rows) and immigration rates (columns). The \arwp\ population (red) and \walbab\ population (blue) lines represent the median with $95\%$ confidence intervals (shaded areas). Black lines represent $10$ random simulation runs. Panel A displays the population over $500$ days under three immigration rates ($0$, $2$, and $10$ mosquitoes per week). The intervention strategies include na\"ive, complete, and maintain. Panel B focuses on the maintain intervention strategy over days 200 to 920 for the same immigration rates. Horizontal dotted lines represent the suppression/establishment ($10\%$) and the unstable equilibrium $\omega^*$ ($40\%$) thresholds used. Values above $1$ on the y-axis indicate population levels higher than the initial steady state. Model parameters are the low values defined in Table~\ref{tab:parameters}.}
    \label{fig:intervention_stormclouds_low}
\end{figure}

\begin{figure}[H]
    \centering
    \includegraphics[width=\linewidth]{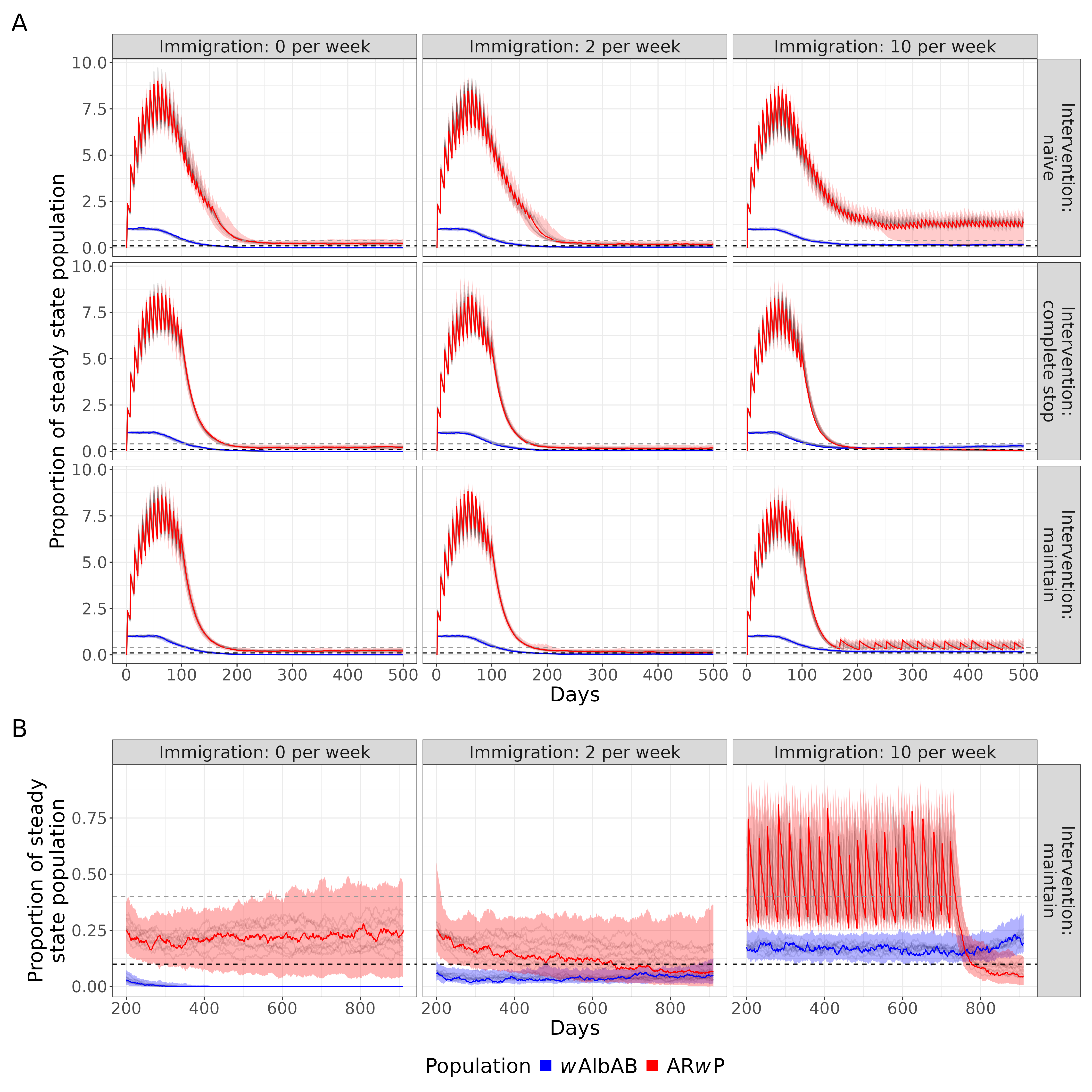}
    \caption{Proportion of mosquitoes relative to the steady-state population over time for different release strategies (rows) and immigration rates (columns). The \arwp\ population (red) and \walbab\ population (blue) lines represent the median with $95\%$ confidence intervals (shaded areas). Black lines represent $10$ random simulation runs. Panel A displays the population over $500$ days under three immigration rates ($0$, $2$, and $10$ mosquitoes per week). The intervention strategies include na\"ive, complete, and maintain. Panel B focuses on the maintain intervention strategy over days 200 to 920 for the same immigration rates. Horizontal dotted lines represent the suppression/establishment ($10\%$) and the unstable equilibrium $\omega^*$ ($40\%$) thresholds used. Values above $1$ on the y-axis indicate population levels higher than the initial steady state. Model parameters are the high values defined in Table~\ref{tab:parameters}.}
    \label{fig:intervention_stormclouds_high}
\end{figure}

\begin{figure}[H]
    \centering
    \includegraphics[width=\linewidth]{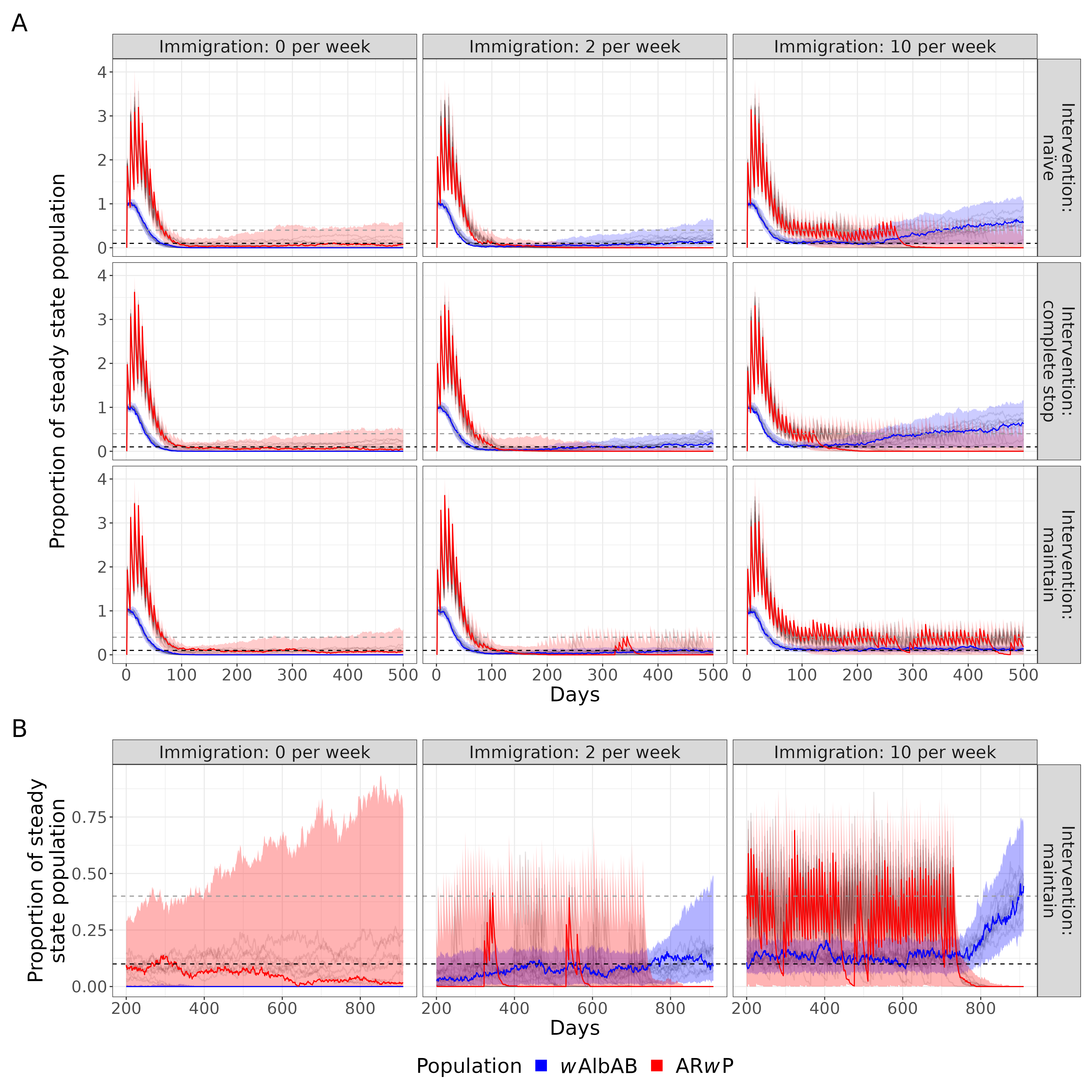} 
    \caption{Proportion of mosquitoes relative to the steady-state population over time for different release scenarios (rows) and immigration rates (columns) and no age-related CI decay. The \arwp\ population (red) and \walbab\ population (blue) lines represent the median with $95\%$ confidence intervals (shaded areas). Black lines represent $10$ random simulation runs. Panel A displays the population over $500$ days under three immigration rates ($0$, $2$, and $10$ mosquitoes per week). The intervention scenarios include na\"ive, complete, and maintain. Panel B focuses on the maintain scenario over days 200 to 920 for the same immigration rates. Horizontal dotted lines represent the suppression/establishment ($10\%$) and the unstable equilibrium $\omega^*$ ($40\%$) thresholds used. Values above $1$ on the y-axis indicate population levels higher than the initial steady state. Model parameters are the expected values  defined in Supplementary Materials Table~\ref{tab:parameters}.}
    \label{fig:intervention_stormclouds_nodecay}
\end{figure}

\begin{table}[H]
      \caption{Percentage of simulations with management success for \walbab\ and \arwp\ populations under different scenarios for immigration rates and intervention strategies. Management success is defined as keeping the \walbab\ population below $10\%$ of the initial population size, and \arwp\ below $40\%$ (unstable equilibrium threshold). The data is shown both within the stopping time and six months after. The model parameters are the low values defined in Table \ref{tab:parameters}.}
      \label{tab:intervention_establishment_low}
      \centering
      \begin{tabular}[t]{lrrrrrr}
      \toprule
      Scenario &  \multicolumn{2}{c}{Na\"ive} & \multicolumn{2}{c}{Complete stop}  & \multicolumn{2}{c}{Maintain}\\
      \cmidrule(lr){2-3} \cmidrule(lr){4-5} \cmidrule(lr){6-7}
      & \walbab & \arwp & \walbab & \arwp & \walbab & \arwp \\
      \midrule
      \addlinespace[0.3em]
      \multicolumn{7}{l}{\textbf{0 per week}}\\
      \rowcolor{gray!6} {\hspace{1em}Within stopping time} & 100 & 100 & 100 & 100 & 100 & 100\\
      \hspace{1em}After six months & 100 & 100 & 100 & 100 & 100 & 100\\
      \addlinespace[0.3em]
      \multicolumn{7}{l}{\textbf{2 per week}}\\
      \rowcolor{gray!6} {\hspace{1em}Within stopping time} & 100 & 100 & 100 & 100 & 91.2 & 100\\
      \hspace{1em}After six months & 50.6 & 100 & 44.6 & 100 & 43.0 & 100\\
      \addlinespace[0.3em]
      \multicolumn{7}{l}{\textbf{10 per week}}\\
      \rowcolor{gray!6} {\hspace{1em}Within stopping time} & 100 & 100 & 100 & 100 & 19.2 & 100\\
      \hspace{1em}After six months & 0.0 & 100 & 0.0 & 100 & 0.0 & 100\\
        \bottomrule
      \end{tabular}
    \end{table} 

\begin{table}[H]
      \caption{Percentage of simulations with management success for \walbab\ and \arwp\ populations under different scenarios for immigration rates and intervention strategies. Management success is defined as keeping the \walbab\ population below $10\%$ of the initial population size, and \arwp\ below $40\%$ (unstable equilibrium threshold). The data is shown both within the stopping time and six months after. The model parameters are the high values defined in Table \ref{tab:parameters}.}
      \label{tab:intervention_establishment_high}
      \centering
      \begin{tabular}[t]{lrrrrrr}
      \toprule
      Scenario &  \multicolumn{2}{c}{Na\"ive} & \multicolumn{2}{c}{Complete stop}  & \multicolumn{2}{c}{Maintain}\\
      \cmidrule(lr){2-3} \cmidrule(lr){4-5} \cmidrule(lr){6-7}
      & \walbab & \arwp & \walbab & \arwp & \walbab & \arwp \\
      \midrule
      \addlinespace[0.3em]
      \multicolumn{7}{l}{\textbf{0 per week}}\\
      \rowcolor{gray!6} {\hspace{1em}Within stopping time} & 100 & 0.0 & 0.0 & 0.0 & 0.0 & 0.0\\
      \hspace{1em}After six months & 100 & 100 & 100 & 100 & 100 & 100\\
      \addlinespace[0.3em]
      \multicolumn{7}{l}{\textbf{2 per week}}\\
      \rowcolor{gray!6} {\hspace{1em}Within stopping time} & 100 & 0.0 & 0.0 & 0.0 & 6.6 & 0.8\\
      \hspace{1em}After six months & 100 & 100 & 100 & 100 & 100 & 100\\
      \addlinespace[0.3em]
      \multicolumn{7}{l}{\textbf{10 per week}}\\
      \rowcolor{gray!6} {\hspace{1em}Within stopping time} & 0.6 & 0.0 & 0.0 & 0.0 & 0.0 & 1.0\\
      \hspace{1em}After six months & 0.0 & 100 & 0.0 & 100 & 0.0 & 100\\
        \bottomrule
      \end{tabular}
\end{table} 

\begin{table}[H]
      \caption{Percentage of simulations with management success for \walbab\ and \arwp\ populations under different scenarios for immigration rates and intervention strategies with no age-related CI decay. Management success is defined as keeping the \walbab\ population below $10\%$ of the initial population size, and \arwp\ below $40\%$ (unstable equilibrium threshold). The data is shown both within the stopping time and six months after. The model parameters are the expected values defined in Table \ref{tab:parameters}.}
      \label{tab:intervention_establishment_Expected_nodecay}
      \centering
      \begin{tabular}[t]{lrrrrrr}
      \toprule
      Scenario &  \multicolumn{2}{c}{Na\"ive} & \multicolumn{2}{c}{Complete stop}  & \multicolumn{2}{c}{Maintain}\\
      \cmidrule(lr){2-3} \cmidrule(lr){4-5} \cmidrule(lr){6-7}
      & \walbab & \wpip & \walbab & \wpip & \walbab & \wpip \\
      \midrule
      \addlinespace[0.3em]
      \multicolumn{7}{l}{\textbf{0 per week}}\\
      \rowcolor{gray!6} {\hspace{1em}Within stopping time} & 100 & 100 & 100 & 100 & 100 & 100\\
      \hspace{1em}After six months & 100 & 99.9 & 100 & 99.9 & 100 & 99.9\\
      \addlinespace[0.3em]
      \multicolumn{7}{l}{\textbf{2 per week}}\\
      \rowcolor{gray!6} {\hspace{1em}Within stopping time} & 100 & 100 & 100 & 100 & 92.6 & 100\\
      \hspace{1em}After six months & 77.9 & 100 & 76.2 & 100 & 49.7 & 100\\
      \addlinespace[0.3em]
      \multicolumn{7}{l}{\textbf{10 per week}}\\
      \rowcolor{gray!6} {\hspace{1em}Within stopping time} & 99.9 & 100 & 99.6 & 99.7 & 23.8 & 100\\
      \hspace{1em}After six months & 0 & 100 & 0 & 100 & 0 & 100\\
        \bottomrule
      \end{tabular}
    \end{table} 

\clearpage
    
\begin{figure}[H]
    \centering
    \includegraphics[width=\textwidth]{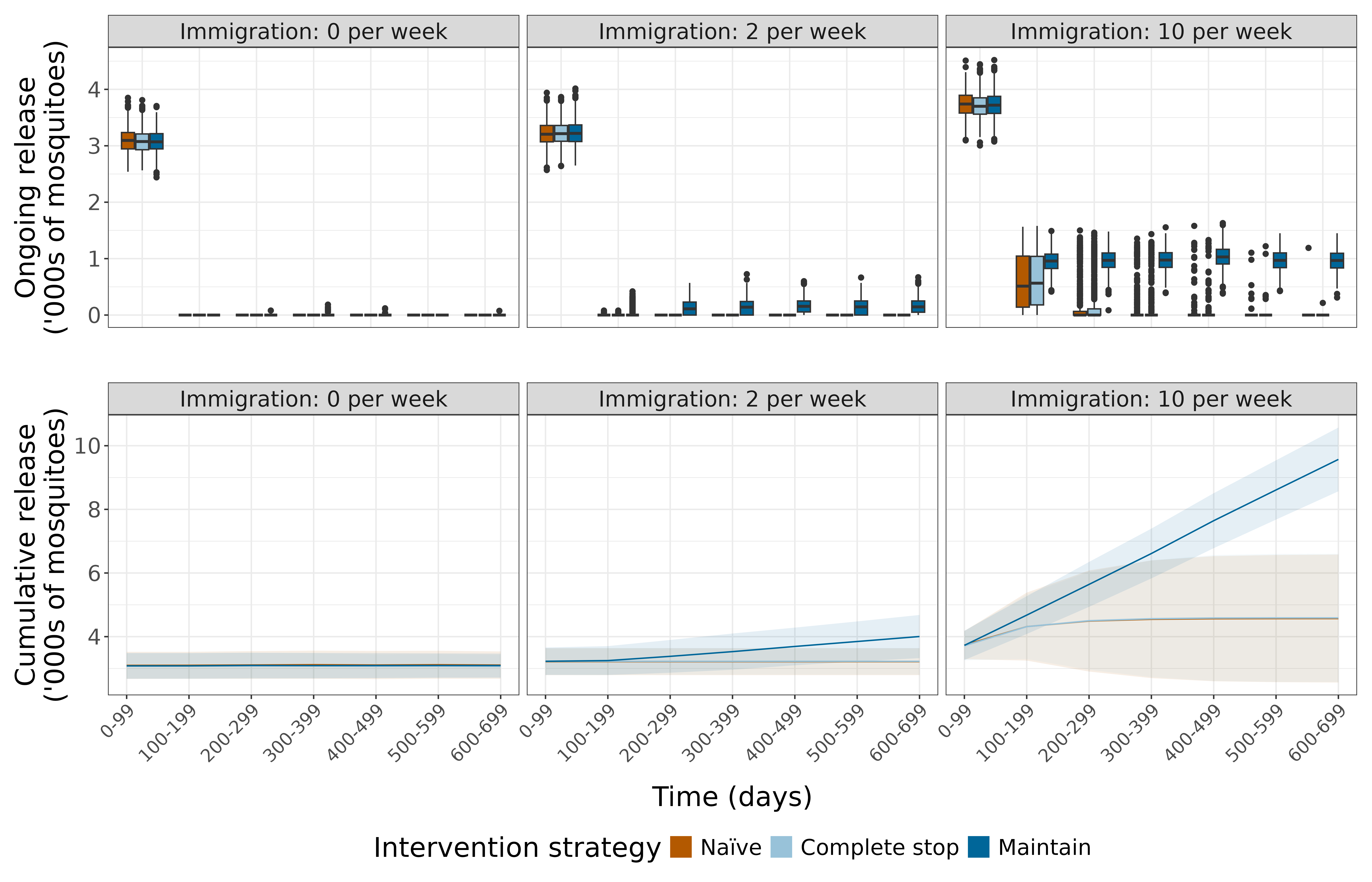}
    \caption{The release in thousands of \arwp\ mosquitoes as a proxy for cost under different intervention strategies and immigration rates. The $x$-axis represents the time in $100$-day intervals up until day $700$. The top row shows the ongoing cost every hundred days, measured as the number of mosquitoes released every $100$ days for each intervention strategy (na\"ive, complete stop, and maintain). The bottom row illustrates the cumulative release (in thousands of mosquitoes) over the entire period for each strategy. The model parameters are the low values defined in Table~\ref{tab:parameters}.}
    \label{fig:intervention_costs_low}
\end{figure}

\begin{figure}[H]
    \centering
    \includegraphics[width=\textwidth]{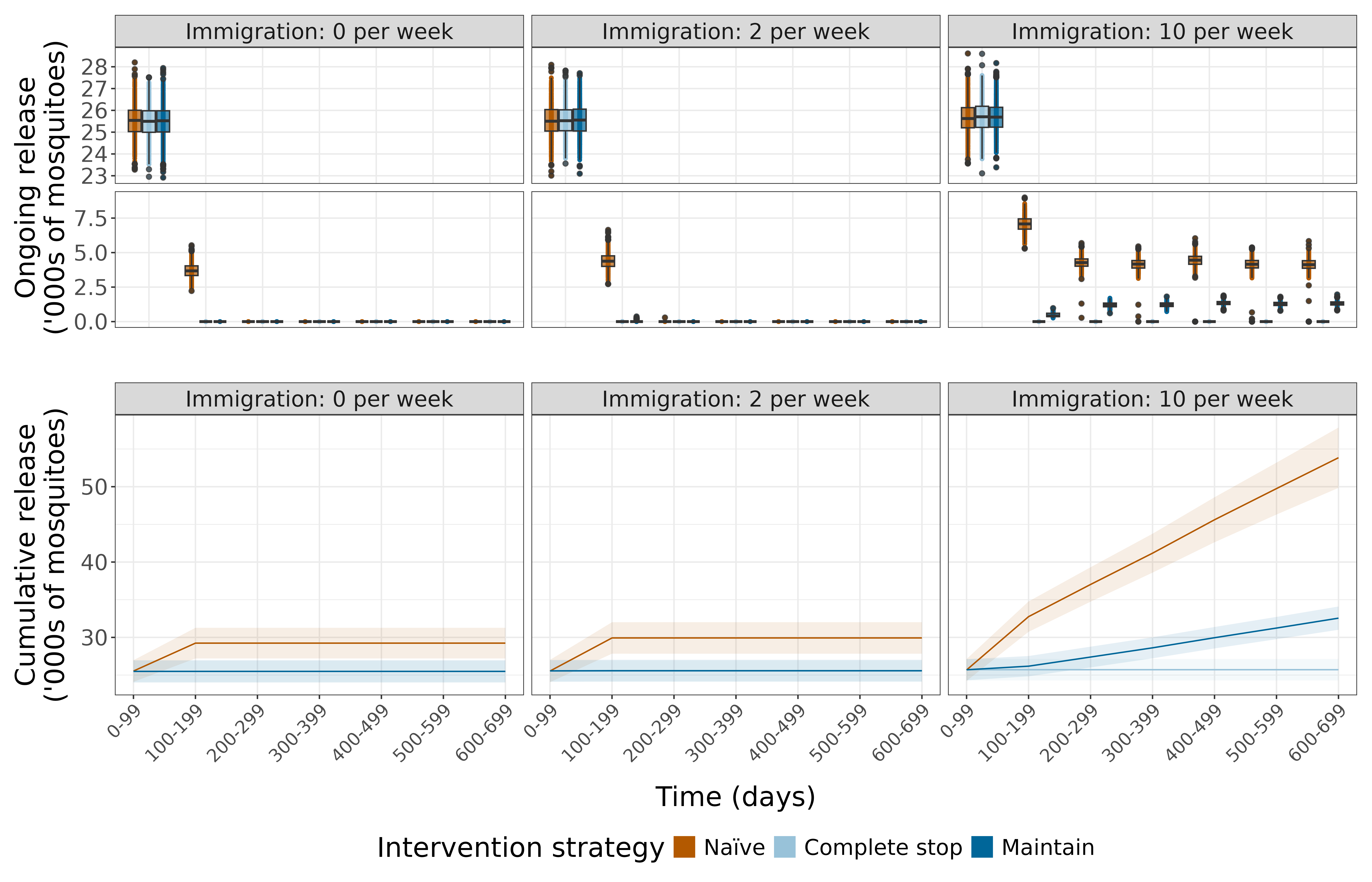}
    \caption{The release in thousands of \arwp\ mosquitoes as a proxy for cost under different intervention strategies and immigration rates. The $x$-axis represents the time in $100$-day intervals up until day $700$. The top row shows the ongoing cost every hundred days, measured as the number of mosquitoes released every $100$ days for each intervention strategy (na\"ive, complete stop, and maintain). The bottom row illustrates the cumulative release (in thousands of mosquitoes) over the entire period for each strategy. The model parameters are the high values defined in Table~\ref{tab:parameters}.}
    \label{fig:intervention_costs_high}
\end{figure}

\begin{figure}[H]
    \centering
    \includegraphics[width=\textwidth]{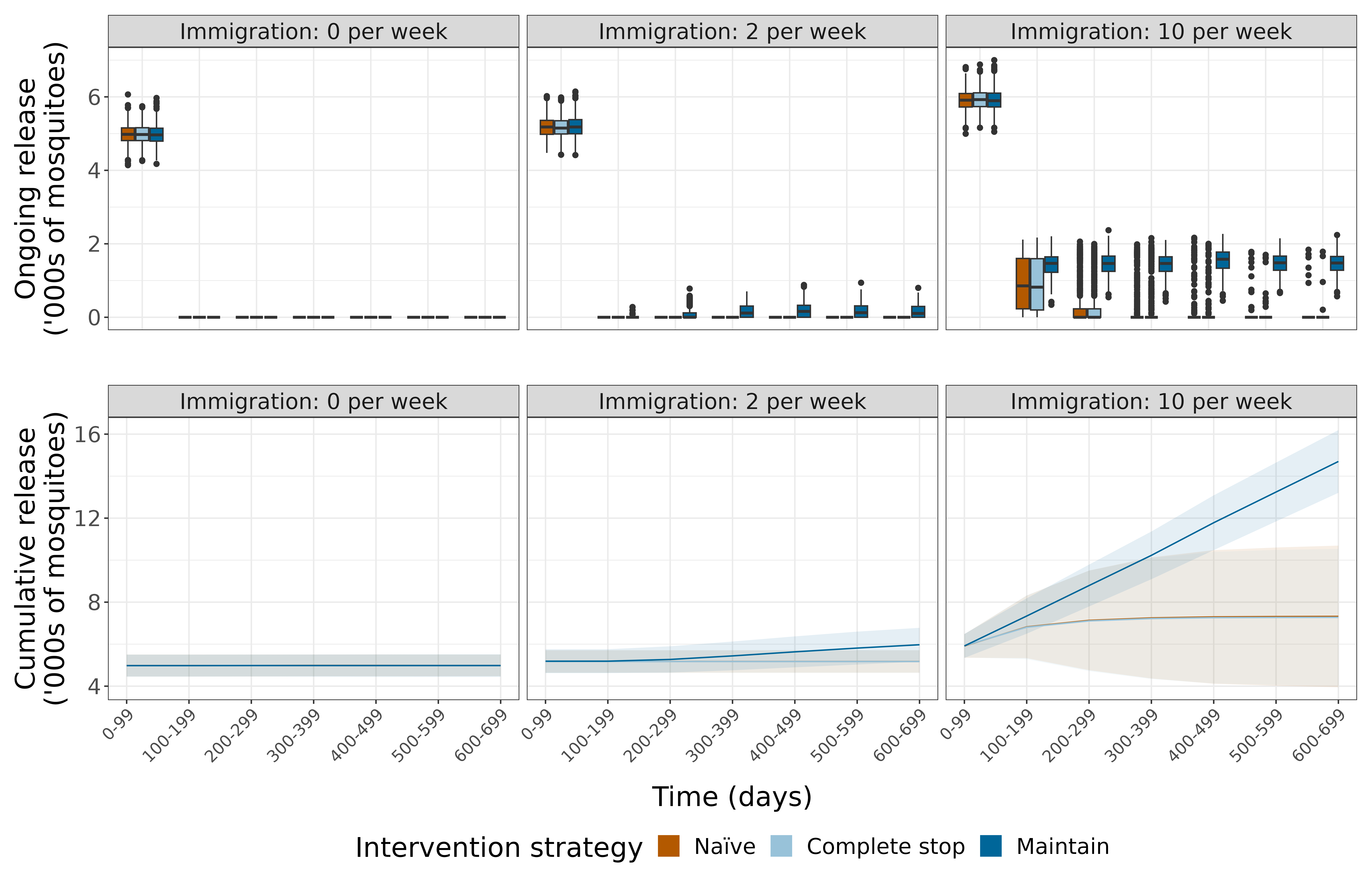}
    \caption{The release in thousands of \arwp\ mosquitoes as a proxy for cost under different intervention strategies and immigration rates with no age-related decay. The $x$-axis represents the time in $100$-day intervals up until day $700$. The top row shows the ongoing cost every hundred days, measured as the number of mosquitoes released every $100$ days for each intervention strategy (na\"ive, complete stop, and maintain). The bottom row illustrates the cumulative release (in thousands of mosquitoes) over the entire period for each strategy. The model parameters are the expected values defined in Table~\ref{tab:parameters}.}
    \label{fig:intervention_costs_nodecay}
\end{figure}

\newpage
\bibliographystyle{plainnat}
\bibliography{refs}